\documentclass[11pt]{article}

\newcommand{\comments}[1]{}

\usepackage[comma,
]{natbib}

\usepackage{overpic}

\usepackage{color,epsfig,eurosym,url,booktabs,soul,enumerate,todonotes,nicefrac}
\usepackage{latexsym}
\usepackage{amsmath}
\usepackage{amssymb}
\usepackage{amsthm}
\usepackage{fullpage,paralist}

\newtheorem{theorem}{Theorem}

\newtheorem{corollary}{Corollary}

\newtheorem{proposition}{Proposition}

\theoremstyle{definition}

\theoremstyle{remark}

\newtheorem{example}{Example}

\newcommand{\footnotes}[1]{\footnote{\renewcommand{\baselinestretch}{1} \footnotesize #1}}

\newcommand{\ta}{\tilde{a}}

\newcommand{\bP}{\mathbf{P}}

\newcommand{\EE}{\mathbb{E}}
\newcommand{\RR}{\mathbb{R}}

\newcommand{\var}{v}

\DeclareMathOperator*{\supp}{supp}
\DeclareMathOperator*{\argmin}{arg\,min}

\DeclareMathOperator*{\tr}{tr}

\newcommand{\DKL}[2]{D\big(#1 \,\big|\big|\, #2\big)}

\begin{document}

\begin{titlepage}

\title{Overconfidence and Prejudice\thanks{We are grateful to Aislinn Bohren, Alex Imas, Robert Lieli, and Florian Zimmermann for insightful discussions, and seminar and conference audiences for comments.}}

\author{Paul Heidhues\\DICE, Heinrich-Heine University D\"{u}sseldorf \and Botond K\H{o}szegi\\Central European University \and Philipp Strack\\Yale University}

\date{\today}

\maketitle

\renewcommand{\baselinestretch}{1.3} \normalsize

\begin{abstract}

We explore conclusions a person draws from observing society when he allows for the possibility that individuals' outcomes are affected by group-level discrimination. Injecting a single non-classical assumption, that the agent is overconfident about himself, we explain key observed patterns in social beliefs, and make a number of additional predictions. First, the agent believes in discrimination against any group he is in more than an outsider does, capturing widely observed self-centered views of discrimination. Second, the more group memberships the agent shares with an individual, the more positively he evaluates the individual. This explains one of the most basic facts about social judgments, in-group bias, as well as ``legitimizing myths'' that justify an arbitrary social hierarchy through the perceived superiority of the privileged group. Third, biases are sensitive to how the agent divides society into groups when evaluating outcomes. This provides a reason why some ethnically charged questions should not be asked, as well as a potential channel for why nation-building policies might be effective. Fourth, giving the agent more accurate information about himself increases all his biases. Fifth, the agent is prone to substitute biases, implying that the introduction of a new outsider group to focus on creates biases against the new group but lowers biases vis a vis other groups. Sixth, there is a tendency for the agent to agree more with those in the same groups. As a microfoundation for our model, we provide an explanation for why an overconfident agent might allow for potential discrimination in evaluating outcomes, even when he initially did not conceive of this possibility.  
\end{abstract}

\comments{maybe think about model in which agent doesn't observe all $c_j$ or all components of $c_j$. Then, can ask questions like: is it a good idea not to report some characteristics of individuals?}

\comments{maybe think about model with omitted variables, in which agent excludes a group with $\theta_k \neq 0$ from his theory.}

\end{titlepage}

\renewcommand{\baselinestretch}{1.5} \normalsize

%
%
%
%
%
%
%
%
%
%
%
%
%
%
%
%
%
%
%
%
%
%
%
%
%
%
%
%
%
%
%
%
%
%
%
%
%
%

\setcounter{page}{1}

\section{Introduction}


Among the many factors that hinder the fair and productive coexistence of different social groups, two sets of widely held beliefs surely stand out. First, many or most people think of their own group as superior to other groups, and often hold negative evaluations of other groups. For instance, the majority of non-Muslim Americans believe that Muslims do not want to adopt the American way of life, a statement that most Muslims disagree with. Second, different groups hold dissenting views about the most urgent intergroup problems. For instance, blacks consider discrimination against blacks a greater problem than whites do --- who often believe that discrimination against whites is going on. Such beliefs can in turn foster or maintain interpersonal discrimination and conflict between groups, especially when resources are scarce \citep{jackson,bertrandduflo}.   

In this paper, we develop a novel theory of prejudiced beliefs based on a person's attempts to understand society while maintaining stubborn, unrealistically positive views of \textit{himself}. Looking to best explain why his outcomes are not as good as he thinks he deserves, the agent comes to overestimate discrimination against any social group he is a member of, and to overestimate discrimination in favor of any group he is a competitor of. Furthermore, interpreting others' outcomes in light of his views about discrimination, he is led to develop excessively positive views of his social groups, and excessively negative views of competing groups. Our framework provides a new perspective on key facts explained by previous theories based on politician-induced hate \citep{glaeser2} and the representativeness heuristic \citep{bordalocoffman}, but we also explain several other facts, including connecting prejudiced beliefs to self-serving beliefs about discrimination, and make a number of further predictions. 

We begin in Section \ref{sec:theoretical_tools} by deriving a general formula for what a person with dogmatic incorrect beliefs about a variable learns about other variables. This formula is invaluable for our subsequent analysis, and might also be useful for studying other learning settings with misspecification and multi-dimensional states and signals. The agent is interested in estimating the levels of $L$ fundamentals, with him having a degenerate prior about one fundamental and a full-support prior over the other fundamentals. He repeatedly observes some linear combinations of the fundamentals with multivariate normal errors that are i.i.d. over time, and updates using Bayes' Rule. We identify his limiting beliefs about the fundamentals and the covariance matrix of the errors, which depend on what linear combinations he observes, the true covariance matrix of the errors, and how wrong his dogmatic belief is. Interestingly, although the agent misinfers the covariance matrix, his inferences about the fundamentals are the same as when he knows the true covariance matrix.  

In Section \ref{sec:model}, we turn to our main topic, a theory of social beliefs. We assume that society is comprised of $K$ potentially overlapping groups, and a person is either a member, a competitor, or a neutral outsider of a group. The agent observes many draws of the ``recognition'' --- i.e., achievement, social status, or other measure of success --- of each individual, including himself. He understands that recognition depends in part on a person's ``caliber'' --- i.e., ability, hard work, or other measure of deservingness --- and noise, but he also posits that there might be ``discrimination'' --- or policies, cheating, conspiracies, etc. --- that benefit a group and hurt its competitors. Some of these possibilities might be realistic, while others might be imagined by the agent; but he does not a priori assume any particular pattern of discrimination, he attempts to learn the true pattern from his observations, including from direct, unbiased signals about the degrees of discrimination themselves. All noise terms in the agent's observations are independent of each other. Crucially, to these relatively standard ingredients we add a single non-classical but empirically well-founded assumption, stubborn overconfidence: the agent has a point belief about his own caliber that is above the correct one. 

In Section \ref{sec:patterns_in_beliefs}, we identify properties of the agent's long-run beliefs, beginning with two central patterns that are widely documented in the literature. First, the agent holds self-serving views about discrimination: he overestimates discrimination against any group he is a member of and underestimates discrimination against any group he is in competition with, and consequently he considers discrimination against his groups as a greater issue than outsiders do. Intuitively, the agent's overconfidence implies that the recognition he obtains is in his view systematically too low, and believing in discrimination against his groups provides an explanation for why this is the case.  

Second, the agent is subject to an ``in-group bias,'' perhaps the most basic finding in the literature on stereotypes, discrimination, and prejudice: he tends to hold overly favorable views about those in the same social circles, and overly unfavorable views about those in competing circles. Intuitively, since the agent believes that discrimination against his groups and in favor of competing groups is going on, he attributes more of an in-group member's recognition, and less of a competitor's recognition, to caliber. 

Beyond naturally accounting for two important stylized facts, our theory makes many subtler predictions. As the intuitions make clear, the degree of the agent's self-serving view of discrimination, and the degree of his in-group bias, are directly related to his degree of overconfidence. Hence, we predict that these biases are increasing in overconfidence, or (equivalently in our model) the extent to which a person feels that he is not getting what he deserves based on his caliber. To our knowledge, this relationship has not been directly tested in the literature, although it is consistent, for instance, with the observation that men are both more overconfident and more prejudiced than women. 

A person's pattern of biases, however, derives not only from his overconfidence, but also from the manner in which he thinks about society. As an illustrative extreme case, suppose that discrimination is in reality non-existent in society. If the agent conceives of each individual separately ($K=0$ groups), then he develops unbiased beliefs about everyone. If he conceives of individuals in terms of group membership ($K>0$ groups), in contrast, then he concludes that groups are being treated differently, and develops in-group biases. This sensitivity of beliefs implies that even asking racist, sexist, or other discriminatory questions, such as ``are Mexicans more likely than Americans to be criminals?'', can be dangerous: if it induces a person to evaluate observations with a new group distinction in mind, then it can lead to the wrong conclusions. Relatedly, our model may contribute to understanding the effects of ``nation-building'' in nineteenth-century Europe and twentieth-century post-colonial Africa. One common element of these efforts was the use of education to encourage thinking in terms of one nation rather than many ethnicities or tribes \citep{miguel,alesinareich}. Our theory says that this can lead different ethnic groups to think more positively of each other. 

While changing the agent's model of society can be helpful, attempts to debias him through the provision of better information may backfire or be only partially effective. An obvious approach would be to target the source of the problem --- overconfidence --- directly by making recognition a less noisy measure of caliber. Ironically, because this forces the agent to provide a better explanation for why his recognition is low, it increases all his biases. Similarly, although better information about discrimination toward a group lowers the agent's bias regarding that parameter, it also creates more of a need to explain his low recognition, increasing many of his other biases. For instance, if someone manages to convince a white male that there is no discrimination against males in hiring decisions, then he will come to believe in discrimination against whites to a greater extent. 

Changes in the environment can also lead to other forms of ``bias substitution'' between groups. For instance, suppose that citizens confronted with the refugee crisis start asking themselves whether immigrants are different or being treated differently. For a new issue such as this one, it is plausible to assume that information about discrimination is poor. Then, citizens come to believe strongly in discrimination toward the new group, and arrive at a strong in-group bias relative to that group. At the same time, their views of other groups improve. Intuitively, the presence of immigrants provides a convenient account for why citizens are not getting what they think they deserve, so they have less need to be prejudiced against other groups. This provides one potential mechanism for how focusing on an outside group can help unify a population hitherto riddled with disagreements and dislikes --- a common tactic of politicians.

The fact that beliefs in our model depend on group membership implies that patterns of agreement and disagreement will often also fall along group lines, at least when it comes to the beliefs of overconfident individuals. Then, mechanisms that have previously been identified in the literature lead to further, potentially detrimental implications. Namely, a person might conclude that outsiders are not only of lower caliber, but also unreasonable or poorly informed, and he might seek out information sources consistent with his group's views. 

In Section \ref{sec:extensions}, we consider variants of our basic model. Most importantly, we ask what happens when the agent does not think in terms of groups ($K=0$), but individuals' recognitions are not necessarily independent of each other. Then, the agent develops a positive bias about any person whose recognition is positively correlated with his, and a negative bias about any person whose recognition is negatively correlated with his. We can then endogenously define the agent's in-group as the former group --- those ``in the same boat'' with him --- and the latter group as his out-group. Furthermore, this case of our model provides a potential microfoundation for why an agent who does not initially think about the possibility of group-level discrimination nevertheless starts believing in it. Namely, comparing his conclusions to the very observations on which his conclusions are based, the agent might observe that his in-group does not get the outcomes that he thinks it should, and draw the natural inference that something must be hurting the entire group. 

We also analyze simple versions of our model in which a person observes not just the recognitions of others, but also signals about their calibers, for instance through personal contact. We show that more precise information about caliber, or knowing a greater number of individuals, lowers all of the person's biases. More precise information about an outside group's recognition, however, lowers the agent's opinion of the group. This insight provides a justification for the convention of news outlets not to report the race of a suspected criminal unless it is essential for the story. Finally, we give an example illustrating that when people are characterized by multiple attributes (e.g., ability and morality), then the agent may develop positive biases about some but not all of a competing group's attributes. 

We discuss related literature throughout, but especially in Section \ref{sec:related_lit}. While a large literature in sociology and social psychology explores prejudices and stereotypes, to our knowledge no previous theory has derived these phenomena from overconfidence or connected them to views about group-level discrimination. As a result, existing theories make some very different predictions than do we. Similarly, Glaeser's \citeyearpar{glaeser2} theory that politicians supply, and citizens often passively accept, hate-creating stories about minorities, and Bordalo et al.'s \citeyearpar{bordalocoffman} theory that individuals exaggerate distinctive true differences across groups, address different aspects of stereotypes and prejudice than our paper.

In Section \ref{sec:conclusion}, we mention some questions that are unaddressed by our current framework. For instance, a person might be unsure about the composition of groups in society. Then, we conjecture that in some circumstances he comes to believe in a secret group that is favored or conspires against others. More broadly, while our framework takes group relationships as exogenous, it would be important to endogenize them in future work.

\section{Theoretical Tools}

\label{sec:theoretical_tools}

In this section, we derive a theoretical result that we will apply in multiple ways to analyze our main models, and that might be useful for other researchers in studying implications of overconfidence and other misspecifications. Readers uninterested in our abstract result can skip to Section \ref{sec:model}.

A person makes inferences about an $L$-dimensional vector of \textsl{fundamentals} 
\[
f = (f_1 , \dots , f_L)^T \in \RR^L\,,
\]
which are fixed over time. In each period $t$, he observes a $D$-dimensional \textsl{signal} 
\[
	r_t = M f + \epsilon_t \,,
\]
where $M \in \RR^{D \times L}$ is a matrix and $\epsilon_t \in \RR^D$ is a vector of errors that is jointly normally distributed with mean zero and positive definite covariance matrix $\Sigma$. We assume that $D\geq L$ and $M$ has rank $L$. Otherwise, there would be two different fundamentals that entail the same distribution of signals and hence the agent could not learn the fundamentals even with access to infinite data.

The agent observes a sequence of realizations of $r_1,r_2,r_3,\ldots$, with the $\epsilon_t$ drawn independently over time. He updates his beliefs in a Bayesian way: given a prior belief $\bP_0$ over the set of fundamentals and positive definite covariance matrices, the probability the agent's posterior belief $\bP_t$ assigns to the set $A$ after seeing the the sequence of signals $r=(r_1,r_2,\ldots,r_t)$ is given by
\[
	\bP_t A = \frac{\int \mathbf{1}_{(f', \Sigma') \in A} \ell_t(r | f',\Sigma') d \bP_0 (f', \Sigma')}{\int  \ell_t(r | f',\Sigma')  d \bP_0 (f', \Sigma')},
\]
where the likelihood equals
\[
	\ell_t (r | f',\Sigma') = \prod_{z = 1}^t \frac{1}{\sqrt{(2 \pi)^{L} \det \Sigma' }} \exp\left(-\frac{1}{2}  (r_z - M f')^T \Sigma' \, (r_z - M f') \right) \,.
\]

Crucially, we assume that the agent is misspecified in a particular sense: while the true value of fundamental $i$ equals $f_i$, he believes with certainty that it is $\tilde f_i$.

We consider three different inference problems depending on what aspect of the agent's beliefs are fixed by his prior belief, and what are derived from his observations. In our preferred specification, the agent is trying to infer the fundamentals $f$ as well as the covariance matrix $\Sigma$:
\begin{equation}\label{eq:no-prior-restrictions}\tag{Case III}
	\supp \bP_0 = \left\{ (f',\Sigma') \in  \RR^L \times \RR^{D \times D} \colon f'_i = \tilde{f}_i, \Sigma' \mbox{ is positive definite} \right\}\,.
\end{equation}
Because they are potentially of interest in other applications, we also consider two simpler inference problems. We ask what the agent infers about the fundamentals when his beliefs about the covariance matrix are fixed at some positive definite $\tilde{\Sigma}$, so that
\begin{equation}\label{eq:prior-fixed-covariance}\tag{Case I}
	\supp \bP_0 = \left\{ (f',\Sigma') \in  \RR^L \times \RR^{D \times D} \colon f'_i = \tilde{f}_i, \Sigma' = \tilde{\Sigma} \right\}\,.
\end{equation}
And we ask what the agent infers about the covariance matrix when his beliefs about \emph{all} fundamentals are fixed at $\tilde{f} = (\tilde{f}_1, \dots ,  \tilde{f}_L)^T$, so that 
\begin{equation}\label{eq:prior-fixed-fundamental}\tag{Case II}
	\supp \bP_0 = \left\{ (f',\Sigma') \in  \RR^L \times \RR^{D \times D} \colon f' = \tilde{f} , \Sigma' \mbox{ is positive definite}\right\}\,.
\end{equation}

We say that the agents' beliefs \emph{concentrate on a point} $(\tilde f, \tilde \Sigma)$ if for every open set $A$ such that $(\tilde f, \tilde \Sigma)$ is contained in $A$, the agent will in the limit assign probability $1$ to $A$:
\[
	\lim_{t \to \infty} \bP_t A = 1 \,.
\]
For stating our theorem, note that any positive definite covariance matrix $\tilde{\Sigma}$ is invertible, so the matrix $M^T \tilde{\Sigma}^{-1} M$ is well-defined; and since $M$ has rank $L$, this matrix is positive definite and hence invertible.  

\begin{theorem}[Long-Run Beliefs] \label{thm:beliefs_no_action} 
In Cases (I), (II), and (III), the agent's beliefs concentrate on a single point $(\tilde f, \tilde \Sigma)$. Furthermore:  
\begin{compactenum}[{Case} (I)]
\item If the agent has fixed beliefs $\tilde{\Sigma}$ about the covariance matrix but is uncertain about the fundamentals $j \neq i$, then in the limit his bias about fundamental $j$ is 
\begin{equation} \label{eq:ability_given_matrix}
	\tilde{f}_j - f_j = \frac{(M^T \tilde{\Sigma}^{-1}  M)^{-1}  _{ij}}{(M^T \tilde{\Sigma}^{-1}  M)^{-1}  _{ii}} (\tilde{f}_i - f_i). 
\end{equation}	
\item If the agent has fixed beliefs $\tilde f$ about the fundamentals but is uncertain about the covariance matrix, then in the limit his bias about the covariance matrix is
\begin{equation}\label{eq:matrix_given_ability}
	\tilde{\Sigma} - \Sigma=  (M (\tilde{f} - f)) (M (\tilde{f} - f))^T .
\end{equation}	

\item If the agent is uncertain about both the fundamentals $j\neq i$ and the covariance matrix, then in the limit his bias about fundamental $j$ is 
\begin{equation} \label{eq:ability_agnostic}
	\tilde{f}_j - f_j = \frac{\left[M^{T} \Sigma^{-1} M\right]^{-1}_{ji}}{\left[M^{T} \Sigma^{-1} M\right]^{-1}_{ii} } \, (\tilde{f}_i - f_i),
\end{equation}    
and his bias about the covariance matrix is given by Expression \eqref{eq:matrix_given_ability}.
\end{compactenum}
\end{theorem}

The proof of Theorem \ref{thm:beliefs_no_action} proceeds as follows. First, by the seminal result of \cite{berk}, beliefs concentrate on the set of minimizers of the Kullback-Leibler divergence. Intuitively, the negative of the Kullback-Leibler divergence is increasing in the subjective likelihood of observing the true distribution of data, so it is a natural measure of how likely a combination of parameters is in the agent view in the long run. Due to our assumption of normal signals, the Kullback-Leibler divergence $\DKL{F,\Sigma}{\hat{f},\hat{\Sigma}}$ assigned to the parameters $(\hat{f},\hat{\Sigma})$ when the true parameters equal $(f,\Sigma)$ simplifies to
\[
	\DKL{f,\Sigma}{\hat{f},\hat{\Sigma}} = \frac{1}{2} \left( \tr(\hat{\Sigma}^{-1} \Sigma ) + (M(\hat f - f))^T \hat{\Sigma}^{-1} M(\hat f - f) - n + \log\frac{\det \hat{\Sigma} }{\det \Sigma}\right)\,.
\]
The proof then derives the unique minimizer of the above expression over the support specified in Cases (I), (II), and (III) using properties of the trace, Kronecker product, determinant, and eigenvalues of a matrix. While Case (I) can be verified by taking first-order conditions with respect to the fundamentals, in Cases (II) and (III) the objective function involves the determinant of $\hat{\Sigma}$, which is not a tractable function in general. We solve this problem by looking at the eigenvalues of a well-chosen matrix in each case, greatly reducing the dimensionality of the problems as well as eliminating the determinant from the objective. 

One curious fact about the agent's inferences is immediate from plugging the true covariance matrix into Expression \eqref{eq:ability_given_matrix} --- which yields exactly Expression \eqref{eq:ability_agnostic}. Hence, Part (III) says that when the agent is initially agnostic about both the fundamentals and the covariance matrix, then --- although by Part (II) he misinfers the covariance matrix --- his beliefs about the fundamentals are the same as if he correctly understood the covariance matrix.

\section{A Model of Inferences about Individuals and Groups} 

\label{sec:model}

We now turn to our main interest, a model of how overconfidence affects social beliefs.

\subsection{Setup}

There are $I$ individuals in society. Individual $j \in \{ 1 , \dots , I \} $ has fixed ``caliber'' $a_j \in \mathbb{R} $ and $K$ observable group relationships $c_j \in \{1,0,-1\}^K$, with $c_{jk} = 1$ denoting that he is a member of group $k$, $c_{jk} = 0$ denoting that he is outside of but neutrally related to group $k$, and $c_{jk} = -1$ denoting that he is in competition with group $k$. We consider society from the perspective of one individual $i\in \{ 1 , \dots , I \} $, whom we call agent $i$; in some cases, we also compare the views of different agents who all think according to our model. Agent $i$ observes a sequence of realizations of
\begin{equation}\label{eq:def performance}
q_j  =  a_j + \sum_{k=1}^K c_{j k} \,\theta_k + \epsilon_j^q\, ,\ j=1,\dots, I, \ \ \  \text{ and } \ \  \ \eta_k  =  \theta_k + \epsilon_k^\eta \, ,\ k=1, \dots, K,	
\end{equation}
where $q_j$ is individual $j$'s ``recognition'' on that occasion, the time-invariant constant $\theta_k \in \RR$ is discrimination in favor of group $k$, $\eta_k$ is a signal of $\theta_k$, and $\epsilon_j^q$ and $\epsilon_k^\eta$ are independent normally distributed errors with mean zero and variances $\var^q_j$ and $\var^\eta_k$, respectively. Hence, recognition depends in part on caliber (and noise), but in addition to that, discrimination toward a group increases a member's recognition by a fixed amount and decreases a competitor's recognition by the same amount.

The crucial assumption of our model --- and the single non-classical assumption from which our results derive --- is that agent $i$ is overconfident about himself. Formally, while his true caliber is $A_i$, he believes with certainty that it is $\ta_i>A_i$. Beyond having an unrealistic self-view, however, agent $i$ is rational: he applies Bayes' Rule correctly to update his beliefs. Furthermore, he is agnostic regarding the levels of discrimination and the calibers of other individuals, with his prior having full support. Similarly, he is uncertain about the covariance matrix of the errors, with a full-support prior over positive definite covariance matrices.

\subsection{Discussion}

Especially given our context of social judgments and prejudice, we think of the variables $a_j$, $q_j$, and $\theta_k$, as well as the concept of overconfidence, quite broadly. To start with the first two, a straightforward interpretation is that $a_j$ is ability and $q_j$ is wage or other measure of economic success that is susceptible to discrimination. Alternatively, $a_j$ could denote a person's deservingness of social rewards based on past work or behavior or general character, with $q_j$ capturing the respect he gets in the form of transfers, perks, or other recognition. For instance, views on whether a low-income person should get transfers often rest on whether the person is seen as hard-working and honest, and attitudes toward immigrants are sometimes framed in terms of who deserves help from the state 
--- with any perceived deviation from fair treatment interpreted as reflecting discrimination. Accordingly, we think of overconfidence as an overly positive view not only of one's ability, but also of one's skills, importance, appropriate status, or deservingness in society. What is crucial for our theory is that the overconfident person thinks of the recognition he obtains as unjustifiably low. And while $\theta_k$ is most straightforwardly interpreted as discrimination, it could also capture policies that favor a group while potentially hurting others, or any actions by group members that favor their own group. For instance, group members may cheat at the expense of outsiders, or they may be part of a conspiracy to benefit themselves. 

Since we assume that recognition in reality as well as recognition as perceived by agent $i$ satisfies Equation \eqref{eq:def performance}, our framework implicitly presumes that agent $i$'s theory includes all groups for which discrimination occurs (i.e., $\theta_k \neq 0$). Our theory does, however, allow for the possibility that some groups in the agent's theory actually face no discrimination ($\theta_k=0$). But even when the agent includes such an ``irrelevant'' group in his model of society, he does not \textit{assume} that there is discrimination in favor of (or against) that group, he merely allows for the possibility that there \textit{might be}, and evaluates what he sees with this possibility in mind. In fact, the agent correctly understands that excluding any group for which $\theta_k \neq 0$ leads to misspecification and therefore incorrect conclusions; and he wrongly believes that if $\theta_k=0$, then he will learn that this is the case. So from his perspective, it is best to include all potentially relevant groups. 

For the purposes of our model, whether a person outside group $k$ is neutral toward or in competition with group $k$ --- i.e., whether $c_{jk}=0$ or $c_{jk}=-1$ --- is based on whether discrimination or other actions that help group $k$ hurt the individual. Technically, we take these relationships as exogenous, although some economic principles might help guide their specification if the agent is reasonable. For example, if a socio-economic group is very small, then the agent may reasonably entertain the idea that there is discrimination in favor of that group, but he cannot reasonably think that such discrimination would hurt him much. It is important to note, however, that the agent may perceive potential competition with groups that are logically not well-founded.

We present the model and results by referring to individual $j$ as a person. In reality, it is unrealistic to assume that a person observes the recognitions of all individuals. An equivalent model obtains if some observations $q_j$ are average recognitions of groups or subgroups. And if agent $i$ observes an individual's or group's recognition with noise, that noise can be incorporated into the error $\epsilon^q_j$.  

The main assumption of our model, that the agent is stubbornly overconfident about himself, is consistent with a large body of evidence from psychology as well as economics that documents overly positive self-views among individuals who have had plenty of opportunity to learn about themselves \cite[see our earlier paper,][for a selective review]{heidhueskoszegistrack}.\footnotes{Following \cite{heidhueskoszegistrack}, we specify agent $i$'s belief about his own caliber as degenerate for two main reasons. First, it allows us to study the implications of overconfidence for inferences in a tractable manner. Second, and more importantly, it is a reduced-form way to incorporate forces modeled in other papers that induce individuals to retain excessively positive self-views. That there are such forces is supported by the plethora of evidence mentioned in our earlier paper. Furthermore, a recent experiment by \cite{goettekozakiewicz} specifically tests and confirms the predictions of our earlier model based on point beliefs.} But beyond this assumption, our model still presumes a lot of rationality in that the agent perfectly applies Bayes' Rule to a diverse set of observations, and develops a single full set of beliefs about all individuals and groups. Indeed, if the agent does not update properly about his caliber, one may wonder whether he makes reasonable inferences about anything. While we cannot think of plausible and tractable specific alternatives, our modeling assumption of Bayesian updating does not seem to drive our intuitive results. The pattern of beliefs the agent develops requires only that he seeks some consistency with his overconfident beliefs, not that he seeks the best possible fit from a Bayesian perspective. And a person need not hold a single set of beliefs about everything. He may, for instance, go through a thought process akin to our updating model every time he is induced to evaluate groups of individuals and discrimination, recalling facts and putting together a story consistent with his overconfidence in a way that depends on the groups he thinks are relevant at the moment. 

Finally, we analyze only the beliefs of individuals, and do not consider the implications of these beliefs for behavior. But a large body of evidence on statistical discrimination reviewed by \cite{bertrandduflo} finds that people act on their beliefs in many social situations, and this has important consequences for economic outcomes. Although most of the literature does not ask whether individuals' beliefs are correct, recent work by \cite{arnolddobbie} on racial bias in bail decisions and by \cite{bohrenimas} on gender bias in performance evaluations shows that the relevant beliefs are biased. This is especially problematic as both theory \citep{coateloury} and evidence \citep{gloverpallais,lavysand,carlana} suggests that stereotypes can become self-fulfilling through the endogenous responses of interacting individuals. We return to this issue briefly in the conclusion.

\section{Patterns in Beliefs}

\label{sec:patterns_in_beliefs}

Theorem \ref{thm:beliefs_no_action} implies that agent $i$'s beliefs converge to point beliefs (in a sense defined precisely in Section \ref{sec:theoretical_tools}). To state our results, we denote agent $i$'s long-run belief about discrimination toward group $k$ by $\tilde{\theta}^i_k$, and his long-run belief about individual $j$'s caliber by $\ta^i_j$. We also denote true discrimination in favor of group $k$ by $\Theta_k$, and individual $j$'s true caliber by $A_j$.

\begin{proposition}[Biases] \label{prop:biases_levels}
Agent $i$'s long-run bias about discrimination toward group $k$ is  
\begin{equation} \label{eq:biases_discrimination}
	\tilde{\theta}^i_k - \Theta_k = \frac{-c_{ik} \var^\eta_k }{\var^q_{i} + \sum_{k'} c_{ik'}^2 \var^\eta_{k'}}\cdot (\tilde{a}_i - A_i),
\end{equation}
and his long-run bias about agent $j$'s caliber is 
\begin{equation}\label{eq:biases_individuals}
	\ta^i_j - A_j = \frac{\sum_k c_{ik}c_{jk}\var^\eta_k }{\var^q_{i} + \sum_k c_{ik}^2 \var^\eta_k}\cdot (\tilde{a}_i - A_i)   .
\end{equation}
\end{proposition}

\noindent Proposition \ref{prop:biases_levels} has a number of important implications for how individuals think about society.

\subsection{Self-Centered Views of Discrimination} 

The first implication is immediately apparent from Equation \eqref{eq:biases_discrimination}: a person overestimates discrimination against any group he is a member of, and underestimates discrimination against any group he is in competition with. Intuitively, overconfidence implies that agent $i$'s average recognition is not on par with his perceived caliber. A compelling explanation is that discrimination against the groups he is in and discrimination in favor of the groups he competes with are hurting him. 

The above bias is difficult to measure directly if the true extent of discrimination is unclear or not easily compared to individuals' opinions. But there is an immediate implication that is amenable to measurement and consistent with evidence: that a member's estimate of discrimination against a group is higher than a non-member's. Regarding racial discrimination, 88 percent of blacks say that ``the country needs to continue making changes to give blacks equal rights with whites,'' while only 54 percent of whites and 69 percent of Hispanics agree with that statement \citep[][Chapter 4]{pew}.\footnotes{More specifically, 70 percent of blacks, but only 37 percent of whites, say that blacks are treated less fairly by police than whites, with similar gaps regarding the treatment of blacks in courts, stores, public schools, health care, and on the job \citep{anderson}.} In fact, the majority of American whites think that they are the ones being discriminated against \citep{nprrwj}. Regarding gender discrimination, 77 percent of surveyed male STEM employees say that women are treated fairly in opportunities for promotion and advancement, but only 43 percent of females agree \citep{funkparker}.\footnotes{Similarly, in a representative survey of Germans between 39 and 50 years of age, 69 percent of women versus 43 percent of men answered that much still needs to be done to accomplish gender-equality (``Weniger Respekt und wachsende Fremdenfeindlichkeit'', Frankfurter Allgemeine Zeitung, 12.09.2019).}
 And in the financial domain, 37\% of those with family incomes over \$75,000, but 56\% of those with family incomes below \$30,000, think that being rich has more to do with having had advantages than with working harder \citep{pew5}.  

It is worth noting that one group the agent conceives of could be a singleton consisting of himself. In this case, he develops the view that there is some ``exclusive'' discrimination directed only at him. If he is in addition particularly bad at judging the degree of exclusive discrimination (the $\Sigma^{\eta}_k$ corresponding to himself is much higher than the $\Sigma^{\eta}_k$ corresponding to other groups), then he converges on what might be called paranoid beliefs: he explains his lack of performance mostly by exclusive discrimination --- the belief that ``the world is out to get him'' and only him. In this case, the group-based biases we identify below are small. But it seems reasonable to posit that the typical person is not so bad at judging the degree of exclusive discrimination (i.e., the $\Sigma^{\eta}_k$ corresponding to himself is not that high). In those cases, all of our conclusions continue to hold.

\subsection{In-Group Bias} 

The agent's biased beliefs about discrimination in turn lead to biased beliefs about individuals. Equation \eqref{eq:biases_individuals} implies that the greater is $\sum_k c_{ik}c_{jk}\var^\eta_k$, the more positively biased is $\ta_j^i$: the more group memberships and the more competing groups individuals $i$ and $j$ have in common, the higher is agent $i$'s opinion of individual $j$. Intuitively, since agent $i$ believes that discrimination against his groups and in favor of competitors is going on, the more groups and competitor groups he shares with individual $j$, the more he believes individual $j$ is suffering from discrimination, so the more of individual $j$'s recognition he attributes to caliber. 

The above bias has two important, closely related implications. For simplicity and clarity, we state these implications in a special case in which a person's in-group and out-groups are unambiguous. We say that the group structure is partitional if the $K$ groups are disjoint, their union is the set of all individuals, and $c_j = c_{j'}$ whenever $j$ and $j'$ are in the same group. This means that society is divided into separate groups, with group memberships determining individuals' relationships to other groups.  

\begin{corollary} \label{cor:in-group_bias}
Suppose that the group structure is partitional, and take two overconfident agents $i_1$ and $i_2$ who belong to different groups $k_1$ and $k_2$.   

1. If the average calibers of the groups are equal, then agent $i_1$ believes that group $k_1$ has higher average caliber than group $k_2$. 

2. Agent $i_1$'s belief about the average caliber of group $k_1$, and his belief about the difference in average calibers between groups $k_1$ and $k_2$, are higher than agent $i_2$'s beliefs about the same measures. 
	
\end{corollary}

Part 1 says that if the average abilities of the groups are (approximately) equal, then all groups believe themselves to be better than other groups. This ``in-group bias'' is perhaps the most basic stylized fact in the literature on stereotypes, discrimination, and prejudice. It was central in the groundbreaking works of \cite{sumner}, \cite{allport} and \cite{tajfel}, and has been confirmed by many researchers \citep[see][for a meta-analysis of 137 studies]{mullenbrown}. As recently as the 90's, for instance, about 65\% of whites expressed the view that whites are more hard-working than blacks, and 55\% thought that whites are more intelligent \citep{krysanmoberg}, while both whites and blacks showed biological prejudice, the most traditional form of prejudiced belief that one's group is innately superior \citep{hrababrinkman}.\footnotes{In the United States, overt expressions of racism have declined over the years; e.g., in a 2014 survey, ``only'' 23\% of whites said that blacks are less intelligent \citep{krysanmoberg}. But evidence suggests that this decline reflects individuals' realization that unambiguous racism is socially unacceptable, and more subtle measures still show substantial prejudice (see for instance \citealt{mcconahayhardee} for an early constribution, and \citealt{fiskenorth} for a review of modern prejudice measures).} Relatedly, \cite{gilens} provides evidence that the widespread dislike of welfare programs in the US is based on the (mis-)perception of whites that recipients are mainly blacks lacking sufficient work ethics, which \cite{alesinaglaeser} argue is an important reason for why the US welfare state is smaller than its European counterparts.\footnotes{Nevertheless, \cite{vanoorschot} finds a similar pattern regarding immigration in Europe: citizens view immigrant groups as less deserving than other needy groups, which he points out is in line with prior research suggesting people close to us in terms of identity are seen as more deserving.} And prejudices can persist even when there is a strong non-discrimination norm: \cite{shayozussman} document that both Jewish and Arab judges favor plaintiffs of their own ethnicity in Israeli small-claims courts, although it is unclear to what extent this reflects beliefs rather than tastes.   

Research on the in-group bias distinguishes between in-group favoritism and out-group derogation \citep{hewstonerubin}, with the former being viewed as a more essential and more common ingredient of in-group bias than the latter. We can define in-group favoritism as the overestimation of in-group members, and out-group derogation as the underestimation of out-group members. Then, our framework says that an overconfident individual always engages in in-group favoritism, but he may or may not engage in out-group derogation. First, if agent $i$ is not a competitor of anyone (for any $k$, $c_{ik} \neq -1$, and for any $j$, $c_{jk^i} \neq -1 $ for $i$'s group $k^i$), then he correctly perceives outsiders. In this case, overestimating discrimination in favor of an outside group does not help agent $i$ in explaining his low performance, so he misestimates only discrimination vis a vis his own group. Since his own group has no competitors, however, this does not affect his view of an outsider. Second, agent $i$'s evaluation of an outsider $j$ is increasing with each competing group they share (i.e., for which $c_{ik} = c_{jk}=-1$), so it can be positively biased. Intuitively, if $i$ and $j$ have a common competitor group, then $i$ believes that $j$ suffers from discrimination in favor of that group just like he does, which introduces a positive bias in his evaluation. Agent $i$ does derogate $j$ if he is, or perceives he is, in competition with $j$ ($c_{jk^i}=-1$ for $i$'s group $k^i$), and they do not have common competing groups --- i.e., they are unequivocal competitors. This pattern is roughly consistent with the basic premise of group conflict theory discussed below, although we account for the same pattern in a different way. 

A specific kind of in-group bias discussed in the literature and predicted by our model is a ``legitimizing myth'' that rationalizes and thereby helps maintain an arbitrary social hierarchy \cite[see social dominance theory, e.g.,][]{prattosidanius}. Suppose that all groups have the same average caliber, but there is a dominant social group that benefits from positive discrimination ($\theta_k >0$), resulting in inequality of outcomes. Then, Corollary \ref{cor:in-group_bias} implies that overconfident members of the privileged group underestimate or completely fail to appreciate the benefits that they are receiving, coming to see inequality as a consequence of real differences between groups. 

An important pattern not fully consistent with our model, however, is that the in-group bias is often stronger for the dominant group than the dominated group, and in extreme cases the dominated group may show a bias toward the dominant group \citep[see][pages 228-234]{sidaniuspratto}. \cite{carddellavigna2} provide a recent example in the academic domain, documenting that both male and female referees appear to be biased toward male authors. Our theory can account for an asymmetric in-group bias by means of additional assumptions, for instance if overconfidence is greater in the dominant group than in the dominated group, but it is inconsistent with a reversed in-group bias. Hence, reasons outside our framework probably contribute to the asymmetry as well. For example, if some members of the dominant group control public discourse, they may present their own interpretations as facts and thereby shift the opinions of all groups.

Part 2 of Corollary \ref{cor:in-group_bias} states variants of the in-group bias when groups' calibers are not necessarily equal, or --- as in the above-mentioned case of a group controlling public discourse --- other forces also act on individuals' views. Even then, an individual evaluates his own group, and his own group relative to other groups, more positively than outsiders do. Evidence is consistent with this implication as well. Regarding ethnic groups, attitudes toward Muslim Americans are significantly more negative among non-Muslims than among Muslims.\footnotes{Specifically, 56 percent of U.S. Muslims, but only 33 percent of the general public, think that Muslims who come to the U.S. want to adopt American customs, and 20 percent of Muslims, but 51 percent of the general public think that Muslims want to be different from the larger American society. Regarding another dimension of the issue, 40 percent of the general public think that there is a great deal or fair amount of support for extremism in the Muslim American community, but only 21 percent of Muslims agree. In reality, Muslim Americans have some distinctly American values and concerns. For instance, 71 percent --- as compared to 62 percent in the general public --- say that most people can get ahead if they are willing to work hard. And Muslims are almost as equally concerned about Islamic extremism as others, with 60 percent very or somewhat concerned about extremism in the U.S., and 72 percent very or somewhat concerned about extremism in the world, versus 67 and 73 percent for the general public. For more details, see the report by the \cite{pew2}.} Regarding gender, while both male and female students give lower evaluations to female instructors than to male instructors, male students do so to a greater extent \citep{mengelsauermann}. And in an international comparison, 88\% of Americans believe that it would be better for the world if the U.S. was the world leader than if China was, but in all other countries surveyed, a lower percentage (from 13\% in Russia to 81\% in Japan) share that view \citep{pew4}.  

Although agent $i$ is biased about in-group members, his bias is limited by his overconfidence: even if he belongs to the exact same groups as individual $j$ ($c_{ik} = c_{jk}$ for all $k$), he has less biased beliefs about agent $j$ than about himself ($\ta^i_j - A_j < \ta_i - A_i$). People are positively biased about those in the same groups, but the median person still believes that he is better than most members of his closest group.

\subsection{Biases Derive from the Perception of Getting Less Than Deserved} 

Another immediate implication of Proposition \ref{prop:biases_levels} is that person $i$'s biases about both individuals and discrimination are increasing in his overconfidence $\tilde{a}_i - A_i$. As we have mentioned, in our model overconfidence has an equivalent interpretation to the belief that one is getting less than what he deserves. Hence, our model predicts that those who feel more strongly that they are getting less than they deserve are more prone to exhibit the biases and prejudices we identify. We have not found direct evidence addressing this prediction but it is consistent with some correlational patterns.\footnotes{While the evidence is not fully conclusive, existing research suggests that males tend to be more overconfident at least regarding ``stereotypical male'' attributes such as mathematical ability \citep{beyer,jakobssonetal,niederlevesterlund}. In as much as males are more overconfident, we predict they have greater belief biases, and some evidence from Sweden and the US suggests that male students indeed have stronger racial prejudices \citep{ekehammarsidanius,quallscox}. Researchers have also studied the relationship between self-esteem and in-group bias, and in line with our prediction have generally found a positive relationship \citep[see, e.g.,][for a review]{abersonhealy}. But we cannot be sure that self-esteem is a good measure of overconfidence, even if it is plausible that they are positively correlated.}

Relatedly, a simple implication of our framework is that a person's biases depend on him trying to explain what is happening to himself. Suppose that in building his theory about society, agent $i$ acted like a disinterested scientist, fitting a model in which he does not treat his own outcomes as observations. Then, he uses a correctly specified model, so he develops correct beliefs about everything. As a specific example, a white male professor studying discrimination in different cities may conclude that racial and gender discrimination are widespread. At the same time, he may be prone to believe that discrimination in his own workplace --- academia --- is non-existent or even that getting hired or promoted is easier for women and minorities.

\subsection{Biases Derive from Group-Based Thinking} 

\label{subsec:group-based thinking}

A person's pattern of biases, however, derives not only from his overconfidence, but also from the manner in which he thinks about society. As a starting point, suppose that $\Theta_k=0$ for all $k$. Then, the biases agent $i$ comes to develop are directly related to the extent to which he evaluates observations with group distinctions in mind.  If he conceives of each individual separately instead of in terms of group membership ($K=0$), then he develops unbiased beliefs about everyone. If he conceives of society in terms of groups ($K>0$), in contrast, he concludes that groups he belongs to or competes with are being treated differently, and develops in-group biases. 

This prediction may contribute to understanding the effects of nation-building efforts that were common in nineteenth-century Europe and twentieth-century post-colonial Africa \citep{miguel,alesinareich}. One common element of nation-building policies was the use of education to encourage thinking in terms of one nation rather than many ethnicities or tribes. Our theory says that this can lead different ethnic groups toward perceiving each other as equals, which --- consistent with evidence by \cite{miguel} --- presumably improves interethnic cooperation. 

To generalize the above insight, we ask what happens when a person adds an irrelevant new group to his conception of society. In any situation (any $K$ and any $\Theta_1, \dots, \Theta_K$), this increases his bias regarding the extent of discrimination:\footnotes{Corollary \ref{cor:group_numbers} is somewhat analogous to Schwartzstein's \citeyearpar{schwartzstein} result that an agent who ignores an important explanatory variable when trying to understand his observations may overestimate the relevance of another variable.}

\begin{corollary}
	Adding an irrelevant new group (group $K+1$ with $\Theta_{K+1}=0$) to the agent's theory increases $\sum_k |\tilde{\theta}^i_k - \Theta_k|$.  
	\label{cor:group_numbers}
\end{corollary}

\noindent If there are more in reality irrelevant groups agent $i$ considers, then he can better explain his observations by developing biased views about these groups, and hence his total bias regarding discrimination increases. As a result, agent $i$'s biases regarding individuals closest to him (for whom $c_{jk}=c_{ik}\neq 0 $ for all $k$) and individuals furthest from him (for whom $c_{ik} =- c_{ik}\neq 0$ for all $k$) also increase. 

At the same time, some of agent $i$'s biases can offset each other, so his bias about a person who shares some but not all group memberships with him can decrease in the number of groups he considers. As a simple example, suppose first that agent $i$ thinks of one group ($K=1$), and $c_{i1} = 1$, $c_{j1} = -1$. Then, agent $i$'s bias about agent $j$ is $\tilde a^i_j - A_j = -\var^\eta_1(\tilde a_i - A_i) /(\var^q_i + \var^\eta_1)$. Now suppose that agent $i$ also considers another division of society ($K=2$), with $c_{i2} = c_{j2} = 1$ and $\var^\eta_1 = \var^\eta_2$. Now agent $i$'s bias is $\tilde a^i_j - A_j = 0$. Intuitively, a white male evaluates a black male more negatively if he thinks of society along a black/white divide than if he thinks of society along black/white as well as male/female divides. 

In some circumstances, however, adding a new group to a person's theory does change his views in a specific direction. In particular, we consider a situation in which agent $i$ starts to contemplate a group of outsiders who are competitors of one of his groups. For instance, men may start asking themselves whether a specific group of women is receiving favorable discrimination at the expense of men. 

\begin{corollary} \label{cor:political correctness}
	Suppose $c_{i\kappa} = 1$, and an irrelevant group $K+1$ satisfying $c_{jK+1} = -1$ for any $j$ with $c_{j\kappa}=1$ is added to agent $i$'s theory. This improves agent $i$'s view of any member of group $\kappa$ and worsens his view of any member of group $K+1$. 
\end{corollary}

\noindent Contemplating the new competitor group, agent $i$ concludes that there is discrimination in favor of it, which lowers his opinion of the new group and improves his opinion of anyone who he believes is hurt by the discrimination. 

Corollary \ref{cor:political correctness} provides one justification for the view --- typically associated with political correctness --- that some questions relating to disadvantaged groups should not be discussed or investigated. The concern behind this view is that groups are innately not different, and many racist, sexist, or other discriminatory questions, such as ``are women as capable as men, or are they getting ahead due to favorable treatment?'' can instead promote prejudiced beliefs. From the perspective of a correctly specified model, such a position is puzzling: if the groups a person investigates are approximately equal, then he will (eventually) conclude that this is the case, lowering any existing prejudices. But in our model, asking such questions is indeed prone to produce prejudiced answers despite equality between groups. Unfortunately, as we have discussed, a person often prefers to ask such questions, and it might be difficult to prevent him.

\subsection{Information about Oneself Increases Biases} 

We identify a few senses in which attempts to address the agent's biases through the provision of information can backfire or be only partially effective. From a classical perspective, the most obvious way to debias a person if his biases derive from an inflated self-view is to provide more accurate information about himself. Indeed, if a person has a correctly specified model with an overly high prior about himself, better information can serve to rectify his self-view faster. But a stubbornly overconfident person's inferences work completely differently: more precise evaluations of himself (i.e., a decrease in $\var^q_{i}$) merely lead agent $i$ to develop stronger biases about everything. Intuitively, being evaluated more clearly forces agent $i$ to acknowledge that his low performance is not due to bad luck, requiring a better explanation for why he is not recognized as he thinks he should be. He responds by increasing his belief about how much discrimination he suffers from, resulting in greater in-group biases as well. This implies that societies or parts of society where evaluations are more frequent or more objectively clear should (all else equal) be more prejudiced.

\subsection{Bias Substitution} 

Attempts to debias can also lead to reallocating a person's biases. As a case in point, suppose that information about discrimination toward one group the agent belongs to or competes with becomes more precise. This could happen due to a social planner providing more information about discrimination, or due to the agent investigating the issue himself. The effect is not fully beneficial: 

\begin{corollary} \label{cor:info about discrimination}
	An increase in the precision of information about discrimination toward a group that agent $i$ either belongs to or competes with (a decrease in $\var^\eta_k$ for a $k$ with $c_{ik}\neq 0$) lowers agent $i$'s bias regarding discrimination toward group $k$ and his total bias regarding discrimination, but increases his bias regarding discrimination toward all other groups.  
\end{corollary}

If agent $i$ receives more information about discrimination toward group $k$, then it becomes more difficult to maintain that discrimination toward group $k$ is going on, and it becomes more difficult to believe in discrimination more generally, so his biases regarding these matters decrease. Looking to explain his recognition in another way, however, agent $i$ engages in bias substitution: he comes to form more biased beliefs about discrimination toward other groups. For instance, if someone convinces a white male that there is no discrimination against males in hiring decisions, then he comes to believe in discrimination against whites to a greater extent.  

Another manifestation of bias substitution occurs when the agent finds a new competitor group to evaluate his observations with, such as when citizens are confronted with the refugee crisis and start asking themselves whether immigrants are different or are being treated differently. For a new issue such as this one, it is plausible to assume that information about discrimination is poor, so that $\Sigma^\eta_k$ is large. 

\begin{corollary}\label{cor:competitor outsiders}
	Suppose that a new competitor group of individuals is added to society (individuals $j=I+1 , \dots , I'$ with $c_{jK+1} =1$ for $j>I$, $c_{jK+1} =-1$ for $j\leq I$, and $c_{jk}=0$ for any $j>I$ and $k\leq K$). Then, agent $i$ develops a negative bias about any member of group $K+1$, and if $\var^{\eta}_{K+1}$ is sufficiently large, then he develops a positive bias about everyone else.     
\end{corollary}

\noindent Intuitively, the presence of immigrants provides a convenient account for why agent $i$ is not getting what he thinks he deserves, so he comes to believe in discrimination in favor of immigrants and develops a negative view of immigrants. But because he views his fellow citizens as also competing with immigrants, he forms positive opinions of them. This provides a mechanism for how focusing on a competitor outside group can help unify a population hitherto riddled with disagreements and dislikes --- a common tactic of politicians. At the same time, agents who do not view themselves as competitors of the new group --- perhaps because they are wealthier and do not compete directly for low-income housing and other state benefits --- do not come to believe in favorable discrimination towards the new group, do not form negative opinions of it, and do not change their opinions of others.

\subsection{Tendency to Agree with In-Group and Disagree with Out-Group} 

The fact that beliefs depend on group membership implies that patterns of agreement and disagreement often also fall along group lines, at least when it comes to the beliefs of overconfident individuals. Suppose that agents $i_1$ and $i_2$ have the same degree of overconfidence: $\tilde{a}_{i_1} - A_{i_1} = \tilde{a}_{i_2} - A_{i_2} $. Then, Equation \eqref{eq:biases_discrimination} implies that agents $i_1$ and $i_2$ agree about the direction of  discrimination toward a group if and only if they have the same relationship with the group ($c_{i_1 k} = c_{i_2 k}$). And Equation \eqref{eq:biases_individuals} implies that agents $i_1$ and $i_2$ agree about the calibers of all individuals if they share all group memberships ($c_{i_1} = c_{i_2}$), and otherwise they may disagree about some or all individuals. Because different biases can cancel each other, however, the degree of agreement is not necessarily decreasing in the number of shared groups.\footnotes{As an example, suppose that $K=2$, $c_{i_1 1} = c_{i_1 2} = 1$, $c_{j1} = 1 , c_{j2}= -1$, and $\Sigma^\eta_1 = \Sigma^\eta_2$. Then, agents $i_1$ and $i_2$ agree about the caliber of individual $j$ if $c_{i_2 1} = c_{i_2 2} = -1$, but they do not agree if $c_{i_2 1} = 1 , c_{i_2 2}= -1$. For instance, a white male and an African American female may evaluate a white female similarly, as they are both biased against her for different reasons. But a white male and a white female do not evaluate the white female similarly.} 

\comments{Botond modified.} Once individuals develop group-based disagreements, mechanisms that have previously been identified in the literature can lead to further, potentially detrimental implications. Analogously to Gentzkow and Shapiro's \citeyearpar{gentzkowshapiro} theory of media bias --- where a consumer rates sources that agree more with his priors to be of higher quality --- the agent finds stories and news reports consistent with discrimination against his in-groups, and with his out-groups being less able, as most credible, and may differentially seek out such sources. Furthermore, an individual's recognition that those outside his circles have different opinions can lead him to conclude that these outsiders are unreasonable or poorly informed, reinforcing the in-group bias we have found above. Worse, just as in Prendergast's \citeyearpar{prendergast2} theory of ``yes-men,'' employees in organizations may --- in an attempt to appear unbiased --- respond by expressing opinions consistent with those in decisionmaking positions. While it applies only to settings where women's actions are observed by men, this insight can help explain the observation that women sometimes display as much gender discrimination as men.\footnotes{\comments{Botond added.} See, e.g., \cite{baguessylos-labini} in the context of academic evaluations, \cite{baguesesteve-volart} in the context of judicial hiring decisions, and \cite{carddellavigna2} in the context of refereeing at top journals. At the same time, other authors, including \cite{zinovyevabagues} studying academic promotions, \cite{gagliarduccipaserman} studying municipal governments, \cite{depaolascoppa} studying academic evaluations, and \cite{kunzemiller} studying promotions at private firms, find that women are treated better by other women than by men.} 

By the logic of our model, the pattern of disagreements above applies only to overconfident individuals. Individuals who have realistic beliefs about themselves also develop realistic beliefs about others and about discrimination. Hence, realistic individuals agree with each other irrespective of group membership, and disagree with overconfident individuals of all groups.  


\section{Model Variants}

\label{sec:extensions}

\subsection{Correlated Errors and Endogenous Groups}

\label{sec:correlated_errors_and_endogenous_groups}

In our main model above, group relationships are exogenous. In the current section, we consider a model in which there are no exogenously given groups that the agent considers relevant to think about, but nevertheless he develops biases. This variant allows us to endogenize a person's in-group and out-group in some situations, and motivates why the agent might want to think of groups and group-level discrimination even in other situations. 

Formally, we make two modifications to our previous model, with all other assumptions remaining unchanged. First, there are no groups, so recognition $q_j = a_j + \epsilon_j$ is an unbiased signal of caliber. Second, the $\epsilon_j$ are not necessarily independent, but have a positive definite covariance matrix $\Sigma^q$. 

A plausible economic example is team production. The $I$ individuals are working in two disjoint teams. Pay is determined by individual performances, which depends in part on individual ability and idiosyncratic noise. But pay also depends on shocks common to the team, such as how well-chosen the tasks are or how a manager evaluates team performance in allocating bonuses. This noise structure induces positive correlation between the outcomes of individuals on the same team, and may induce negative correlation between the outcomes of individuals on different teams. 

Biases are now determined in the following way:

\begin{proposition}[Correlated Errors and Biases] \label{prop:correlated_errors} Agent $i$'s long-run bias about agent $j$ is 
\begin{equation} 
	\ta^i_j - A_j  =  \frac{\Sigma^q_{ij}}{\Sigma^q_{ii}} (\ta_i - A_i) ,	
\end{equation}    
while his bias about the covariance matrix is given by
\begin{equation} 
		 \tilde{\Sigma}^q_{jj'} - \Sigma^q_{jj'} = (\ta^i_j - A_j)(\ta^i_{j'} - A_{j'})= \frac{\Sigma^q_{j'i} \Sigma^q_{ji} }{{\Sigma^q_{ii}}^2}  (\ta_i - A_i)^2  \,.
\end{equation}
\end{proposition}

To start developing intuition for Proposition \ref{prop:correlated_errors}, suppose first that agent $i$ has a correct understanding of the covariance structure of signals. As before, the basic implication of agent $i$'s overconfidence is that he repeatedly observes levels of $q_i$ that seem to him surprisingly low. If he knows that $q_i$ and $q_j$ are positively correlated, then his conclusion that $q_i$ is systematically too low leads him to conclude that $q_j$ must be systematically too low as well. As a result, he overestimates individual $j$.  

By the second part of Proposition \ref{prop:correlated_errors}, however, the agent misestimates the covariance matrix as well; specifically, he overestimates the covariance between $q_j$ and $q_{j'}$ if and only if he misestimates individuals $j$ and $j'$ in the same direction. For an intuition, suppose that he overestimates both individuals. Then, in a prototypical observation both $q_j$ and $q_{j'}$ seem to him to be unexpectedly low and thus positively correlated.  

Finally, Proposition \ref{prop:correlated_errors} implies that agent $i$'s misestimation of the covariance matrix does not affect his inferences about individuals. Intuitively, the amount by which agent $i$ overestimates $a_j$ relative to $a_i$ (i.e., $(\ta^i_j - A_j) / (\ta_i - A_i)$) both determines the relative amount by which he overestimates the covariance of $q_j$ and $q_i$ ($(\tilde{\Sigma}^q_{ij} - \Sigma^q_{ij})/(\tilde{\Sigma}^q_{ii} - \Sigma^q_{ii})$), and is determined by his relative estimate of that covariance ($\tilde{\Sigma}^q_{ij}/\tilde{\Sigma}^q_{ii}$). This can only be consistent if he estimates the relative covariance ($\tilde{\Sigma}^q_{ij}/\tilde{\Sigma}^q_{ii}$) correctly.\footnotes{Formally, Part II of Theorem \ref{thm:beliefs_no_action} implies that $\nicefrac{(\tilde{\Sigma}^q_{ij} - \Sigma^q_{ij})}{(\tilde{\Sigma}^q_{ii} - \Sigma^q_{ii})} = \nicefrac{(\ta^i_j - A_j)}{(\ta_i - A_i)}$; and Part I of Theorem \ref{thm:beliefs_no_action} implies that $\nicefrac{(\ta^i_j - A_j)}{(\ta_i - A_i)} = \nicefrac{\tilde{\Sigma}^q_{ij}}{\tilde{\Sigma}^q_{ii}}$. For both equations to hold simultaneously, it must be that $ \nicefrac{(\tilde{\Sigma}^q_{ij} - \Sigma^q_{ij})}{(\tilde{\Sigma}^q_{ii} - \Sigma^q_{ii})} = \nicefrac{\tilde{\Sigma}^q_{ij}}{\tilde{\Sigma}^q_{ii}}.$ Dividing by the right hand side and rewriting yields $\nicefrac{(1- \Sigma^q_{ij}/\tilde{\Sigma}^q_{ij})}{( 1- \Sigma^q_{ii}/\tilde{\Sigma}^q_{ii})} =1$, implying that $ \nicefrac{\tilde{\Sigma}^q_{ij} }{\tilde{\Sigma}^q_{ii} } = \nicefrac{ \Sigma^q_{ij}}{\Sigma^q_{ii} }$.} 

While agent $i$'s long-run conclusions depend on the correlation structure of the signals, a realistic person's long-run conclusions do not. For an overconfident agent to make correct inferences, individuals must not only be evaluated in an unbiased way, they must be evaluated in an independent way from him.

This model allows us to endogenize a person's in-group as individuals whose outcomes are positively correlated with his --- those ``in the same boat'' with him --- and his out-group as individuals whose outcomes are negatively correlated with his. With this endogenous specification of the in-group and out-group, the model predicts the same type of in-group bias that we have identified in Section \ref{sec:patterns_in_beliefs}, as well as an interesting pattern of biases in agent $i$'s perception of the covariance structure. Specifically, agent $i$ overestimates the covariance between the outcomes of two in-group members as well as the covariance between the outcomes of two out-group members, but he underestimates the covariance between the outcomes of an in-group member and an out-group member. These biases are consistent with one aspect of perceived group homogeneity defined by \cite{linvillefischer2} and documented by \cite{quattronejones}, that a person overestimates how much one group member's outcome predicts another's outcome. Perceived group homogeneity is an interesting contrast to correlation neglect, whereby people perceive or assume less correlation between relevant variables than there is in reality \citep{demarzovayanos,eysterrabin,enkezimmermann}.   

Furthermore, the current model suggests one possible explanation for why individuals might want to estimate theories of discrimination, as we have assumed exogenously in Sections \ref{sec:model} and \ref{sec:patterns_in_beliefs}. Comparing  his conclusions to the very performances that generated his conclusions, agent $i$ might notice that his in-group is faring persistently worse, and his out-group is faring persistently better, than he thinks they should given their calibers. Even if he initially did not think so, he might begin to suspect that discrimination is going on. As a result, he might be drawn to evaluate the data allowing for discrimination.  

A natural question is what happens when --- combining our model in Section \ref{sec:model} with that here --- agent $i$ allows for group-level discrimination, and individuals' recognitions are correlated. In fact, our proof of Proposition \ref{prop:biases_levels} allows for this possibility, and implies that the effects are additive: agent $i$'s opinion of individual $j$ is increasing both in the covariance between their recognitions and in the number of groups and competing groups they share.

\subsection{Personal Contact}


In this section, we consider how an agent's inferences are modified if he also observes signals of the calibers of individuals. This could happen, for instance, if he has personal contact with or a trustworthy source about some members of society, so that he receives signals about them that are not tainted by discrimination.  

We assume that agent $i$ makes the same observations as in the model of Section \ref{sec:model}, and also observes signals of individuals' calibers. Because the general analysis appears intractable, however, we solve a special case of the model. We assume that there is only one group, and each individual is either a member or a competitor of the group ($c_j \in \{ -1 , 1\}$); we drop the subscript $1$ for the single group.  Furthermore, recognitions $q_j = a_j + c_j \theta + \epsilon^q_j$, signals about caliber $s_j= a_j + \epsilon^a_j$, and signals about discrimination $\eta= \theta + \epsilon^\eta$ are independently distributed with $\epsilon^q_j \sim N(0,v^q), \epsilon^a_j \sim N(0,v^a)$, and $\epsilon^\eta \sim N(0,v^\eta)$. This means that individuals' recognitions have the same variance, and so do signals about individuals' calibers.  


\begin{proposition}[The Effect of Personal Contact]\label{prop:general_richer_obs}
Agent $i$'s long-run bias about discrimination is 
$$
\tilde{\theta}^i - \Theta = \frac{-v^{\eta}(v^q+v^a)c_i}{(v^q +v^{\eta})(v^q+v^a) + (I-1)v^qv^{\eta} }\cdot (\tilde{a}_i - A_i),
$$	
and his long-run bias about individual $j$ is 
$$
\tilde{a}^i_j - A_j = \frac{v^{\eta} v^a c_i c_j}{(v^q +v^{\eta})(v^q+v^a) + (I-1)v^qv^{\eta}}\cdot (\tilde{a}_i - A_i).
$$	
\end{proposition}  

The qualitative pattern of biases is similar to that before: the agent is prone to believe in discrimination against his group and in favor of the out-group, and he develops positive biases regarding his in-group and negative biases regarding his out-group. But Proposition \ref{prop:general_richer_obs} also implies that more accurate information about individuals' calibers (a lower $v^a$) and observing more people (a higher $I$) both lower all biases. Intuitively, the former makes it more difficult to maintain one's biases about individuals, and the latter provides better information about the role of discrimination in performances by allowing a person to compare those performances to his direct observations about caliber. For instance, getting to know many members of an out-group might make it clear that their achievements are not due to favorable discrimination.  

In the comparative static above, a reduction in $v^a$ applies to observations of both in-group members and out-group members. Although the intuition applies generally, unfortunately we cannot solve a model in which the variances are different. To help confirm the effect of improved information about just out-group members, we consider a particular example.  

\begin{example}\label{prop:richer_observations}
Suppose that $I=4$, with individuals 1 and 2 being members and individuals 3 and 4 being competitors of the group. Agent 1 observes an out-group member's recognition with variance $v^q_{o}$, and an out-group member's caliber with variance $v^a_{o}$. The variances of all other errors are 1. Then, 
\begin{equation}\label{eq:richer_observations_abilities}
	\frac{\ta^1_3 - A_3}{\tilde{a}_1 - A_1} = \frac{\ta^1_4 - A_4}{\tilde{a}_1 - A_1} = \frac{- 2 v^a_{o} }{5v^q_{o} + 5 v^a_{o} + 4 } \ \ \text{and} \ \   
	\frac{\tilde{\theta}^1_1 - \Theta_1}{\tilde{a}_1 - A_1} = \frac{-2 (v^q_{o} + v^a_{o}) }{5v^q_{o} + 5 v^a_{o} + 4}.
\end{equation}	
\end{example}

\noindent Confirming the previous logic, better information about the out-group's caliber (a reduction in $v^a_o$) lowers all of the agent's biases. This makes the agent more realistic not only about his out-group and the extent of discrimination, but also about his in-group. Intuitively, receiving more accurate information about the out-group's caliber makes it more difficult to believe that the group's average caliber is low, which in turn makes it more difficult to believe that discrimination in favor of the out-group is going on. As a result, it is also less viable to believe that the in-group has high caliber.

The prediction of our model that contact between different groups can reduce prejudices and biases is an instance of Allport's \citeyearpar{allport} influential contact hypothesis, for which the evidence is overwhelming. \cite{pettigrewtropp} provide a meta-analysis of hundreds of studies, most of which find evidence consistent with the hypothesis. Many studies are correlational in nature, which are suggestive albeit not well-identified.\footnotes{As a simple illustrative example from a poll by the \cite{pew3}, 52 percent of Americans agreed with the statement that immigrants are a burden because they take jobs and housing. It is unlikely that the same percentage of immigrants agree, so this is probably another instance of in-group bias. But more important for the present purpose is the geographic variation in these beliefs. In areas with a high concentration of foreign-born individuals, only 47 percent of those with U.S.-born parents think that immigrants are a burden, whereas in areas with a low concentration of foreigners, 65 percent think so. Of course, an alternative interpretation is that immigrants settle in places that are friendlier toward them.} But evidence reviewed by \cite{paluckgreen} in which researchers experimentally manipulate interactions between groups shows that contact has a causal negative impact on prejudices.  

Proposition \ref{prop:general_richer_obs} also implies that observing more accurate information about individuals' \textit{recognitions} rather than caliber is detrimental. Part of the reason is the same as in our basic model: better information about himself makes it more imperative for the agent to explain why his recognition is low, increasing all his biases. But Example \ref{prop:richer_observations} makes it clear that there is another effect acting through observations of the out-group. Observing more accurate information about the recognition of the out-group (i.e., a decrease in $v^q_{o}$) decreases the agent's misinference about discrimination, but it increases his bias about the out-group's caliber. The intuition is the following. Given that agent 1 overestimates discrimination in favor of the out-group, the actual recognition of the out-group is worse than he expects. He attributes this difference to noise, but a decrease in the noise makes such an attribution less plausible. As a result, he concludes that discrimination must not be as strong, but also that the out-group must be of lower caliber, than he thought. For example, many majority Eastern Europeans believe that the Roma receive positive discrimination from police and get away with crimes too easily. Our model says that providing more information about how badly the Roma are treated will lead the majority to conclude not only that favorable treatment is not as pronounced, but also that the Roma are committing worse crimes, than they previously thought.  

This prediction of our model provides a potential justification for the practice of mainstream news outlets not to report the race of a suspected criminal unless it is essential for the story.\footnotes{See for example Guideline 12.1 of the German Press Codex (available at \url{https://www.presserat.de/fileadmin/user_upload/Downloads_Dateien/Pressekodex13english_web.pdf}).} Under the assumption that all parties use a correctly specified model, it is difficult to understand this practice, especially when in reality groups' crime rates do not differ much. But in our model, giving more precise information about outsiders' outcomes --- as would be the case if racial information was provided --- creates or exacerbates incorrect, prejudiced views, so it can be seen as harmful. It is important to emphasize, however, that other misspecifications on the part of readers or journalists can also render racial information in crime reporting detrimental.\footnotes{For instance, some journalists may hold prejudiced views that affect when and how they report a criminal's race. If readers do not account for journalists' prejudices when interpreting reports, they can develop biased views. And within our framework, reporting racial information can also encourage readers to think in terms of races and the moral standing of different races. If so, our model says that this can increase racial biases (Section \ref{subsec:group-based thinking}).}

\subsection{Multi-Dimensional Attributes}

\label{subsec:multidimensional}

In our models above, each individual is characterized by a single attribute $a_i$. In reality, people think of others in multidimensional ways. In this section, we consider a simple example of such richer conceptualizations of individuals; a thorough analysis is beyond the scope of the paper.  

\comments{Botond added brief proof in appendix.}

\begin{example} \label{ex:multi-dim_attr} There is one group and two individuals, with individual 1 being a representative member and individual 2 being a representative competitor of the group. Agent 1 makes observations about the social statuses of his in-group and out-group, which equal
\begin{align*}
	q_1 & = a_1 + m_1 + \theta_1 + \epsilon^q_1  \text{ and} \\
	q_2 & = a_2 + m_2 - \theta_1 + \epsilon^q_2 ,
\end{align*}
where $a_j$ is group $j$'s talent, $m_j$ is group $j$'s morality, and $\theta_1$ is discrimination in favor of the in-group. Agent 1 also observes his out-group's business success, which equals
$$
b_2 = 2 a_2 + m_2 + \epsilon^b_2.
$$ 
Hence, business success is unaffected by discrimination, and depends relatively more on talent than does social status. Finally, the agent observes a signal of discrimination $\eta_1  = \theta_1 + \epsilon^\eta_1$ as before. We assume that the agent is overconfident regarding his total deservingness of social status, overestimating $a_1+ m_1$ by $\Delta_1$. The errors are independent, with the variance of $\epsilon^q_1$ being $v^q_1$ and the variance of $\epsilon^\eta_1$ being $v^\eta_1$. Then, agent 1 develops the following biases in the long run:
$$
\ta^1_2 - A_2 = \frac{1}{1+ v^q_1/v^\eta_1}\cdot \Delta_1 ; \ \ \tilde{m}_1 - M_1 = \frac{-2}{1+ v^q_1/v^\eta_1}\cdot \Delta_1 ; \ \ \tilde{\theta}_1 - \Theta_1 = \frac{-1}{1+ v^q_1/v^\eta_1} \cdot \Delta_1 .
$$
\end{example}

As in our previous models, the agent comes to believe that discrimination against his in-group and in favor of the out-group is going on. But he does not develop exclusively negative views of the out-group as a result: he comes to think that the out-group in more talented than it really is. Intuitively, given that he believes discrimination in favor of the out-group is going on, he underestimates the out-group's total deservingness of status, $a_2 + m_2$. At the same time, he must reconcile his beliefs with realistic views of the out-group's business success, which are more sensitive to $a_2$. The best way to do so is to slightly overestimate the out-group's talent and grossly underestimate its morality.  

Of course, the above overestimation depends on the specific pattern of observations the agent makes, and is therefore a possibility result rather than a general prediction of our model. Nevertheless, it is consistent with the observation that stereotypes about out-groups are sometimes positive. For instance, Jews used to be stereotyped as smart and hard-working, women are often seen as being kind and empathic, and some minority men are considered good athletes. At first sight, this may seem to contradict the idea that individuals tend to hold negative or at best realistic views of their out-groups. Yet exactly as in our example, even when some stereotypes are positive, they often go along with, or even form part of, broader and arguably more important negative stereotypes (see \citealt{jackson}, pp. 18-20, for a discussion, and \citealt{glickfiske0} and \citealt{fiskecuddy0} for closely related ideas). That Jews are smart and hard-working goes along with the idea that they are cool and competitive. That women are kind goes along with the view that they are not capable leaders. And that minorities are good athletes goes along with the notion that they are not good at academics.

\section{Related Literature}

\label{sec:related_lit}

In this section, we relate our theory to research not discussed elsewhere in the paper. 

Because the agent draws conclusions from observations while holding an incorrect view about himself, conceptually our paper belongs to the growing literature on learning with misspecified models. Researchers have studied inferences by individuals who ignore some explanatory variables \citep{remamullainathan,schwartzstein}, misunderstand causal relationships \citep{spiegler8}, misinterpret social observations \citep{bohren,bohrenhauser}, are overconfident \citep{leyaouanqnestermann,heidhueskoszegistrack}, or make mistakes in applying Bayes' Rule \citep[e.g.,][]{rabinschrag,rabinlsn}. The specific economic questions we ask and the specific theoretical methods we use are different from those in the literature. 

The predominant economic approach to stereotypes --- i.e., generalizations about groups --- is that of statistical discrimination, in which individuals use available information correctly to make inferences about individuals \citep{phelps,arrowdiscrimination}. In our setting, the agent also uses his observations to form beliefs, but he does so incorrectly. In this sense, our model can be thought of as one of misspecified statistical discrimination. 

From a psychological perspective, our model is most closely related to social identity theory \citep{tajfelturner,tajfel}. Social identity theory posits that individuals see themselves as members of relevant social groups --- their in-groups --- and identify with those groups. As a result, their self-esteem is bound up with their in-groups, so thinking positively about their in-groups and negatively about their out-groups leads them to think and feel positively about themselves. Our theory also implies that a person's prejudices are intimately tied to his views about himself, but the connection follows a different --- in a sense reverse --- logic: a person thinks positively about himself, and this leads to biases about his in-groups and out-groups. By virtue of being the most consistent beliefs with an inflated self-view, the prejudices in our model can also be interpreted as helping to maintain a high self-esteem.

Relatedly, group conflict theory posits that competition between two groups for the same limited resource naturally leads to hostility between the groups, as well as discrimination and prejudice \citep{jackson}. We derive prejudiced beliefs from intrapersonal considerations.\footnotes{See also \cite{akerlofkranton} for a theory about the effects of identity on behavior, which are less related to our theory focused on beliefs.}

Another strand of the social psychology literature conceptualizes stereotypes as heuristic simplifications of real attributes of groups. \cite{bordalocoffman} formalize this idea using a version of Kahneman and Tversky's \citeyearpar{kahnemantversky4} representativeness heuristic. They assume that a person considers a trait more typical in a group if it is relatively more common in the group than in the relevant comparison group. This approach does not comfortably explain why stereotypes are often derogatory prejudices and why many views are self-serving, and unless different groups have different comparison groups, it also does not explain why different groups hold different views. On the other hand, our framework does not explain neutral stereotypes, such as the view that Swedes are blonde, which the framework of Bordalo et al. does.  

\cite{glaeser2} presents a political-economy model of hate, which he defines as beliefs about the harmfulness of others. Politicians can send fake messages that the out-group is dangerous, and these messages are costly for the electorate to investigate. Because voters who believe that the out-group is dangerous prefer policies that lower the out-group's resources, politicians benefit from hate-inducing messages that complement their policies. For instance, a pro-redistribution politician might want to induce hate against rich minorities. Unlike our framework, this model explains how the political environment can affect people's beliefs about minorities, and which messages are communicated by which politicians. At the same time, our theory helps understand why negative attitudes often persist without politicians stoking them, or even despite politicians' attempts to debias. In addition, our theory can be viewed as one explanation for Glaeser's \citeyearpar{glaeser2} assumption that only hateful messages can be sent: negative messages about other groups resonate more with citizens because these fit better with pre-existing beliefs.

\section{Conclusion}

\label{sec:conclusion}

Our theory posits exogenously given groups that are known to individuals. What happens when --- as in reality --- groups are endogenous and not fully known is an interesting question for future research. As a simple illustration, consider a young academic who is unsure about what determines publication success but knows that he is not a member of a privileged group that accepts each other's papers at the expense of others. As he observes that his papers do not get the credit he overconfidently believes they deserve, he concludes that there must be such a group. As a result, he tries to find the group and become a member of it (rather than improving papers). Since he never finds the group, he develops the conspiracy theory that it must be a secret society.

While we study only beliefs in this paper, ultimately we are interested in how actions interact with biased beliefs. Among the many possible questions, consider the troubling finding in the literature on stereotypes we have mentioned: that biased beliefs can become self-fulfilling through a variety of mechanisms. Our model provides a platform for exploring such mechanisms. For instance, researchers have found that a salient negative stereotype (a ``stereotype threat'') can directly affect a stereotyped person's performance in the relevant domain \citep{steelearonson}. This changes the observations people make about the person for the worse, exacerbating the stereotype and potentially creating a vicious circle that ends in a real performance gap far greater than the bias itself.

\renewcommand{\baselinestretch}{1} \normalsize

\bibliographystyle{aer}
\bibliography{../biblio/economics}

\renewcommand{\baselinestretch}{1.5} \normalsize

\begin{appendix}

\section{Proofs}

For brevity, throughout the Appendix we denote the bias of the agent's long-run beliefs about fundamental $j$ by
\[
\Delta_j = \tilde f_j - f_j ,
\]
and let $\Delta =(\Delta_1 , \dots , \Delta_L)^T$. \\

\noindent \textbf{Proof of Theorem \ref{thm:beliefs_no_action}.} 
As shown in \citep[main theorem p.54]{berk} the support of the agent's beliefs will concentrate on the set of points that minimize the Kullback-Leibler divergence to the true model parameters ($f,\Sigma$) over the support of $\bP_0$
\begin{equation}\label{eq:DKL-minimizer}
	\argmin_{(\hat{f},\hat{\Sigma}) \in \supp \bP_0 } \DKL{f,\Sigma}{\hat{f},\hat{\Sigma}} ,
\end{equation}
where the Kullback-Leibler divergence is given by
\[
	\DKL{f,\Sigma}{\hat{f},\hat{\Sigma}} = \EE\left[ \log \frac{\ell_1 (r_1 | f,\Sigma) }{\ell_1 (r_1 | \hat{f},\hat{\Sigma})}\right]\,.
\]
We will argue that the minimization problem \eqref{eq:DKL-minimizer} admits a unique solution when the prior $\bP_0$ satisfies either \eqref{eq:prior-fixed-covariance}, \eqref{eq:prior-fixed-fundamental}, or \eqref{eq:no-prior-restrictions} and thus beliefs concentrate on a single point.
As both the true model as well as the subjective model are Normal, we have that the Kullback-Leibler divergence simplifies to\footnotes{See for example \url{https://en.wikipedia.org/wiki/Kullback\%E2\%80\%93Leibler_divergence#Multivariate_normal_distributions}}
\begin{equation}\label{eq:DKL-Normal}
	\DKL{f,\Sigma}{\hat{f},\hat{\Sigma}} = \frac{1}{2} \left( \tr(\hat{\Sigma}^{-1} \Sigma ) + (M(\hat f - f))^T \hat{\Sigma}^{-1} M(\hat f - f) - D + \log\frac{\det \hat{\Sigma} }{\det \Sigma}\right)\,.
\end{equation}
Throughout, we denote by $\tilde{f},\tilde{\Sigma}$ the agents subjective long-run beliefs about the mean of the fundamentals and the covariance matrix.
Define the matrix 
\[
B=M^T\tilde{\Sigma}^{-1}M \in \RR^{L \times L}
\] and denote it's elements by $(B_{jk})_{j,k \in \{1,\ldots,L\}}$. For future reference, note that since $\tilde{\Sigma}$ is symmetric, so is $M^T\tilde{\Sigma}^{-1}M$, and thus $B_{jk}=B_{kj}$. Furthermore, as $\tilde{\Sigma}$ is positive definite, so is $\tilde{\Sigma}^{-1}$ and $B=M^T \tilde{\Sigma}^{-1} M$.\medskip\\
We first analyze \underline{Case (I)}: By condition \eqref{eq:prior-fixed-covariance} the minimum in \eqref{eq:DKL-minimizer} is taken over means of the fundamentals $\hat{f}$ or equivalently biases $\Delta = \hat{f} - f$, taking the subjective covariance matrix $\hat{\Sigma}=\tilde{\Sigma}$ as given. By Berk's Theorem, the agent's beliefs about the fundamentals concentrate on the set that minimizes the Kullback-Leibler divergence \eqref{eq:DKL-Normal}. As we can ignore all terms that do not depend on $\hat{f}$, we get that the support of the subjective long-run belief about the mean of the fundamental is contained in
\begin{align}
	\argmin_{\hat{f} \colon\hat{f}_i = \tilde{f}_i} (M(\hat f - f))^T \tilde{\Sigma}^{-1} M(\hat f - f) &= f + \argmin_{\Delta \colon \Delta_i= \tilde{f}_i-f_i} \Delta^T \left( M^T \tilde{\Sigma}^{-1} M \right) \Delta \nonumber \\
	&= f+\argmin_{\Delta \colon \Delta_i= \tilde{f}_i-f_i} \,\sum_{k = 1}^L \sum_{j = 1}^L B_{k j} \Delta_k \Delta_j \label{eq:Objective-Delta}\,.
\end{align}
Here the sum symbolizes the addition of $f$ to every element by element in the set of minimizers.
Taking the first order conditions in the bias about fundamental $\Delta_h$ for $h\neq i$ and using that $B_{jk}=B_{kj}$ yields 
\[
0= 2 \sum_{k=1}^L B_{k j} \Delta_k\,.
\]
Dividing by 2 and plugging in $\Delta_k = \frac{B^{-1}_{k i}}{B^{-1}_{i i}} \Delta_i$ on the right-hand-side yields
\begin{align*}
	\sum_{k=1}^L B_{k j} \Delta_k  = \sum_{k=1}^L B_{k j} \frac{B^{-1}_{k i}}{B^{-1}_{i i}} \Delta_i = \frac{\Delta_i}{B^{-1}_{i i}}  \sum_{k=1}^L B_{k j} B^{-1}_{k i} 
	= \frac{\Delta_i}{B^{-1}_{i i}}  \sum_{k=1}^L B_{j k} B^{-1}_{k i}
	=\frac{\Delta_i}{B^{-1}_{i i}} (B B^{-1})_{j i} \,,
\end{align*}
which equals zero as  $B B^{-1}$ is the identity and $i \neq j$. Hence, $\Delta_k = \frac{B^{-1}_{k i}}{B^{-1}_{i i}} \Delta_i$ satisfies the first order condition.

Let $e_k$ be the $k$-th unit vector, for $k \in \{1,\ldots,L\}$.  We next verify that the first order condition is sufficient for a global minimum. To do so, we rewrite the part of the objective \eqref{eq:Objective-Delta} in terms of $\Delta_{-i} = \sum_{j \neq i} e_j \Delta_j$
\begin{align}
	\Delta^T B \Delta &= \left(e_i \Delta_i + \sum_{j \neq i} e_j \Delta_j \right)^T B \left(e_i \Delta_i + \sum_{j \neq i} e_j\Delta_j \right) 
	= \left(e_i \Delta_i + \Delta_{-i} \right)^T B \left(e_i \Delta_i + \Delta_{-i} \right)\nonumber \\
	&= \left(e_i \Delta_i  \right)^T B \left(e_i \Delta_i  \right) + \Delta_{-i}^T B \Delta_{-i} + 2 \left(e_i \Delta_i\right)^T B \Delta_{-i} \label{eq:objDelta} \,.
\end{align}
The Hessian with respect to $\Delta_{-i}$ of \eqref{eq:objDelta} equals $2 B$. As any quadratic form with a positive definite matrix Hessian has a unique global minimum that satisfies the first-order condition, it follows that indeed 
\[
\Delta_k = \frac{B_{k i}^{-1}}{B_{i i}^{-1}} \Delta_i = \frac{(M^T \tilde{\Sigma}^{-1}  M)^{-1}  _{ij}}{(M^T \tilde{\Sigma}^{-1}  M)^{-1}  _{ii}} \Delta_i
\]
is the unique global minimizer for all $k \neq i$. This completes (I).\medskip\\

We next analyze \underline{Case (II)}: In this case the agent takes the subjective mean of the fundamentals $\tilde f$ and thus the bias $\Delta$ as given and estimates the covariance matrix $\tilde{\Sigma}$. Again, by Berk's Theorem the agent's beliefs about the covariance matrix concentrate on the set that minimizes the Kullback-Leibler divergence \eqref{eq:DKL-Normal}, which is equivalent to the set 
\begin{equation}\label{eq:objective-given-fundamental}
	\argmin_{\hat{\Sigma}}  \left( \tr(\hat{\Sigma}^{-1} \Sigma ) + (M \Delta)^T \hat{\Sigma}^{-1} (M \Delta)  + \log\frac{\det \hat{\Sigma} }{\det \Sigma}\right)\,.
\end{equation}
Denote by $\cdot \otimes \cdot:\RR^D \times \RR^D \to \RR^{D \times D}$ the Kronecker product. In matrix notation, we want to show that the unique minimum of \eqref{eq:objective-given-fundamental} is attained at
\[
\hat{\Sigma} = \Sigma + (M \Delta) \otimes (M \Delta)^T
\]
To simplify notation let $y = M \Delta$.
We first manipulate the objective function 
\begin{align}
	 \tr&(\hat{\Sigma}^{-1} \Sigma ) + y^T \hat{\Sigma}^{-1} y + \log\frac{\det \hat{\Sigma} }{\det \Sigma} 
	 = \tr(\hat{\Sigma}^{-1} \Sigma ) + \tr(y^T \hat{\Sigma}^{-1} y) + \log( \det \hat{\Sigma}) - \log (\det \Sigma)   \nonumber\\
	& = \tr(\hat{\Sigma}^{-1} \Sigma ) + \tr(\hat{\Sigma}^{-1} [y \otimes y^T] ) - \log( \det \hat{\Sigma}^{-1}) - \log (\det \Sigma)   \nonumber \\
	& = \tr\Big(\hat{\Sigma}^{-1} (\Sigma + [y \otimes y^T] )\Big) - \log\Big( \det \hat{\Sigma}^{-1}\Big) - \log (\det \Sigma) \nonumber \\
	&= \tr\Big(\hat{\Sigma}^{-1} (\Sigma + [y \otimes y^T] )\Big) - \log \det \Big(\hat{\Sigma}^{-1} (\Sigma + [y \otimes y^T] )\Big) + \log \det \Big( \Sigma^{-1} (\Sigma +[y \otimes y^T] ) \Big) \nonumber \\
	&= \tr\Big(\hat{\Sigma}^{-1} (\Sigma + [y \otimes y^T] )\Big) - \log \det \Big(\hat{\Sigma}^{-1} (\Sigma + [y \otimes y^T] )\Big) + \log \det \Big(  Id + \Sigma^{-1}[y \otimes y^T] \Big)  \, . \label{eq:DKL-transformed}
\end{align}
Here we used in the first equality that a real number equals it's trace and the log of the ratio equals the difference of the logs. The second equality uses that the trace of $A^T B$ equals the trace of $B A^T$. For third equality we use that the trace is an additive function.
In the last equalities we use  that the sum of logarithms equals the logarithm of the product and that the product of determinants equals the determinant of the product.
Now notice that since $\Sigma$ and $y$ do not depend on $\hat{\Sigma}$, the set of minimizers equals
\begin{align}
	 \argmin_{\hat{\Sigma}} \,\,\, \tr(\hat{\Sigma}^{-1} (\Sigma + [y \otimes y^T] )) - \log( \det (\hat{\Sigma}^{-1}(\Sigma + [y \otimes y^T]) ). \label{eq:transformed-min}
\end{align}
Let $\lambda_1, \dots , \lambda_D$ be the eigenvalues of the matrix $\hat{\Sigma}^{-1} (\Sigma + [y \otimes y^T] )$. Since the trace is the sum of eigenvalues  and the determinant is the product of eigenvalues, \eqref{eq:transformed-min} is minimized by all matrices $\hat{\Sigma}$ such that the eigenvalues of $\hat{\Sigma}^{-1} (\Sigma + [y \otimes y^T])$ minimize
\begin{equation}\label{eq:eigenvalue-characteriztion}
	\sum_{k=1}^D \lambda_k - \sum_{k=1}^D \log \lambda_k. 
\end{equation}
As \eqref{eq:eigenvalue-characteriztion} is strictly convex, we can take the first order condition to identify the unique minimizer. This yields that \eqref{eq:eigenvalue-characteriztion} uniquely minimized if and only if  $\lambda_k = 1$ for all $k$. As all eigenvalues equal one and $\tilde{\Sigma}^{-1} (\Sigma + [y \otimes y^T])$ is symmetric---and hence diagonalizable---, $\tilde{\Sigma}^{-1} (\Sigma +[y \otimes y^T] )$ is the identity matrix. This establishes that
\begin{equation}\label{eq:hatSigma}
	\tilde{\Sigma} = \Sigma + [y \otimes y^T] = \Sigma + (M \Delta) \otimes (M \Delta)^T
\end{equation}
is the unique minimizer of \eqref{eq:objective-given-fundamental} and thus the subjective long-run belief of the agent about the covariance matrix. This establishes (II).\\\medskip

Finally, we prove \underline{Case (III)}: Again, by Berk's Theorem the agent's long-run bias about the fundamental and beliefs about the covariance matrix concentrate on the set that minimizes the Kullback-Leibler divergence \eqref{eq:DKL-Normal} 
\begin{equation}\label{eq:objective-given-both}
	\argmin_{(\Delta,\hat{\Sigma})\colon \Delta_i = \tilde{f}_i - f_i}  \frac{1}{2} \left( \tr(\hat{\Sigma}^{-1} \Sigma ) + y^T \hat{\Sigma}^{-1} y  - D + \log\frac{\det \hat{\Sigma} }{\det \Sigma}\right)\,.
\end{equation}
As shown in \eqref{eq:DKL-transformed} this objective is equivalent to $\nicefrac{1}{2}$ times
\[
	\tr\Big(\hat{\Sigma}^{-1} (\Sigma + [y \otimes y^T] )\Big) - \log \det \Big(\hat{\Sigma}^{-1} (\Sigma + [y \otimes y^T])\Big) - D + \log \det \Big(  Id + \Sigma^{-1}[y \otimes y^T] \Big) \,.
\]
Plugging in the minimizer for the covariance matrix $\Sigma + [y \otimes y^T]$ derived in part two simplifies the objective to
\begin{equation}
	\log \det \Big(  Id + \Sigma^{-1}[y \otimes y^T] \Big) \,.\label{eq:simplified-objective} 
\end{equation}
We first observe that as the determinant is the product of eigenvalues, \eqref{eq:simplified-objective} equals the sum of the logarithms of the eigenvalues of $Id + \Sigma^{-1}[y \otimes y^T]$. Furthermore, if $\lambda$ is an eigenvalue of $Id + \Sigma^{-1}[y \otimes y^T]$ with associated eigenvector $v$ then $\lambda-1$ is an eigenvalue of $\Sigma^{-1}[y \otimes y^T]$ as
\[
	\lambda v = (Id + \Sigma^{-1}[y \otimes y^T]) v \Rightarrow (\lambda-1) v = \Sigma^{-1}[y \otimes y^T] v \,.
\]
If we denote the eigenvalues of $\Sigma^{-1}[y \otimes y^T]$ by $\lambda_1,\ldots,\lambda_D$ then the objective \eqref{eq:simplified-objective} equals
\[
	\sum_{i=1}^K \log(\lambda_k+1)\,.
\]
As eigenvalues are independent of the basis, we next choose an orthogonal basis $x_1,\ldots,x_D$ such that $x_1 = y$ (we can always do so by picking an arbitrary basis and applying the Gram-Schmidt process). Denote, $\mathbf{1} = (1)$ the $1\times 1$ identity matrix.  As $x_i$ is orthogonal to $y=x_1$, we have that
\[
	\Sigma^{-1}[y \otimes y^T] x_i = \Sigma^{-1}[y \otimes y^T] [ \mathbf{1} \otimes x_i ] =  \Sigma^{-1}[y \mathbf{1}] \otimes [ y^T  x_i ] = \begin{cases}
0 &\text{ if } i \neq 1\\
(y^T y) (\Sigma^{-1} y)  &\text{ if } i =1
\end{cases} \,.
\]
Hence, $D-1$ of the eigenvalues of $\Sigma^{-1}[y \otimes y^T]$ equal zero. We will next show that $v = \Sigma^{-1} y$ is an eigenvector with associated non-zero eigenvalue. Let $v = \sum_{i=1}^D \alpha_i x_i$ be the representation of $v = \Sigma^{-1} y$ in the basis $x$. We have that
\[
	\Sigma^{-1}[y \otimes y^T] v = \alpha_1 (y^T y) (\Sigma^{-1} y) = \alpha_1 (y^T y) v
\]
and thus $v$ is an eigenvector of $\Sigma^{-1}[y \otimes y^T]$ with eigenvalue $\alpha_1 (y^T y)$. As $\alpha_1$ is given by the projection of $v$ on $y$, we have that $\alpha_1 = \frac{y^T v}{y^T y}$ and thus the non-zero eigenvalue of $\Sigma^{-1}[y \otimes y^T]$ equals 
\[
	\alpha_1 (y^T y)=  y^T v = y^T \Sigma^{-1} y \,.
\]
Consequently, the agents long-run belief about the mean of the state satisfies
\begin{align*}
	\tilde{f} &= f + \argmin_{\Delta \colon \Delta_i= \tilde{f}_i-f_i} y^T \Sigma^{-1} y\\
	&= f + \argmin_{\Delta \colon \Delta_i= \tilde{f}_i-f_i} \Delta^T \left( M^T \Sigma^{-1} M \right) \Delta\,.
\end{align*}
By (I) we have then have that the unique minimizer and thus the long-run belief of the agent is given by
\begin{align}\label{eq:solution}
	\begin{aligned}
	\Delta_k &= \frac{\left[M^{^T} \Sigma^{-1} M\right]^{-1}_{ki}}{\left[M^{^T} \Sigma^{-1} M\right]^{-1}_{ii} } \Delta_i\, && \text{ for } k \neq i \\
	\tilde{\Sigma} &= \Sigma + (M {\Delta}) \otimes (M {\Delta})^T 
	\end{aligned} \,.
\end{align}
This completes the proof of (III).\qed

\bigskip

\noindent \textbf{Proof of Proposition \ref{prop:biases_levels}.} Let $\Sigma^q,\Sigma^\eta$ be the variance-covariance matrices of $\epsilon^q$ and $\epsilon^{\eta}$,
\begin{align*}
	\Sigma^q &= \text{diag}(\var_1^q, \ldots, \var_I^q)\\
	\Sigma^\eta &= \text{diag}(\var_1^\eta,\ldots, \var_K^\eta)
\end{align*}
 and observe that they are invertible as the variances are greater than zero. We show that this model can be reduced into our old model. To see this observe that one can write the vector $(q \,\eta)^T$ in matrix notation as
\begin{equation}
\begin{pmatrix}
	q\\
	\eta\\
\end{pmatrix} = \begin{pmatrix}
	Id & C\\
	0 & Id\\
\end{pmatrix}\cdot
\begin{pmatrix}
	a\\
	\theta\\
\end{pmatrix}+
\begin{pmatrix}
	\epsilon^q\\
	\epsilon^\eta\\
\end{pmatrix}\,.
\end{equation}
Let $$M=\begin{pmatrix}
	Id & C\\
	0 & Id\\
\end{pmatrix}.$$ As $M$ has determinant $1$ it is invertible.
We have that the matrix $\left[M^{^T} \Sigma^{-1} M\right]^{-1}$ is given by
\begin{align*}
\left[M^{^T} \Sigma^{-1} M\right]^{-1} &= M^{-1} \Sigma (M^{-1})^T = \begin{pmatrix}
	Id & -C\\
	0 & Id\\
\end{pmatrix} \begin{pmatrix}
	\Sigma^q & 0\\
	0 & \Sigma^\eta \\
	\end{pmatrix}\begin{pmatrix}
	Id & 0\\
	-C^T & Id\\
\end{pmatrix} \\
&=\begin{pmatrix}
	Id & -C\\
	0 & Id\\
\end{pmatrix} \begin{pmatrix}
	\Sigma^q & 0\\
	- \Sigma^\eta C^T & \Sigma^\eta \\
	\end{pmatrix}= \begin{pmatrix}
	\Sigma^q + C \, \Sigma^\eta \, C^T & - C \Sigma^\eta \\
	- \Sigma^\eta C^T & \Sigma^\eta \\
	\end{pmatrix} \,.
\end{align*}
By Theorem \ref{thm:beliefs_no_action} agent $i$'s bias about the ability of agent $j$ is given by
\begin{align*}
	\ta^i_j - A_j &= \frac{\left[M^{^T} \Sigma^{-1} M\right]^{-1}_{ij}}{\left[M^{^T} \Sigma^{-1} M\right]^{-1}_{ii}} \Delta_i
	= \frac{\left[\Sigma^q + C \, \Sigma^\eta \, C^T\right]_{ij}}{\left[\Sigma^q + C \, \Sigma^\eta \, C^T\right]_{ii}} (\tilde{a}_i - A_i) \\
	&= \frac{\sum_k c_{ik}c_{jk}\var^\eta_k }{\var^q_{i} + \sum_k c_{ik}^2 \var^\eta_k}\cdot (\tilde{a}_i - A_i)  \,.
\end{align*}
By a similar argument we have that the estimated bias associated with characteristic $k$ is given by
\begin{align*} 
	\tilde{\theta}^i_k - \Theta_k &=  \frac{\left[M^{^T} \Sigma^{-1} M\right]^{-1}_{i(I+k)}}{\left[M^{^T} \Sigma^{-1} M\right]^{-1}_{ii}} \Delta_i
	= \frac{\left[- \Sigma^\eta C^T\right]_{i k}}{\left[\Sigma^q + C \, \Sigma^\eta \, C^T\right]_{ii}} (\tilde{a}_i - A_i) \\
	&=
	\frac{-c_{ik} \var^\eta_k }{\var^q_{i} + \sum_k \sum_k c_{ik}^2 \var^\eta_k}\cdot (\tilde{a}_i - A_i).
\end{align*}
This proves the result.\qed

\bigskip

\noindent \textbf{Proof of Corollary \ref{cor:in-group_bias}.} \noindent Part 1. Consider individual $i$, who is a member of group $k$. For any individual $j$ in group $k$, $c_{ik} c_{jk}=1$, so by Equation \eqref{eq:biases_individuals} individual $i$ overestimates individual $j$. Hence, individual $i$ overestimates the average ability of group $k$. For any $k'\neq k$ and member $j$ of group $k'$, we have $c_{ik'} c_{jk'}\leq 0$, so individual $i$ does not overestimate individual $j$. As a result, individual $i$ does not overestimate the average ability of group $k'$. 

Given that the average abilities of the groups are equal and $i_1$ overestimates the average caliber of $k_1$ but not of $k_2$, the result follows. 

\noindent Part 2. Since, by the reasoning in the first paragraph of the proof of Part 1, $i_1$ overestimates the ability of group $k_1$ but $i_2$ does not,  $i_1$ believes the average ability of $k_1$ to be greater than $i_2$ does. And because in addition $i_2$ overestimates the average ability of $k_2$ while $i_1$ does not, $i_1$ thinks that $k_1-k_2$ is greater than $i_2$ does. \qed

\bigskip

\noindent \textbf{Proof of Corollary \ref{cor:group_numbers}.} Using Equation \eqref{eq:biases_discrimination}, we have 
$$
\sum_k |\tilde{\theta}^i_k - \Theta_k|  = \frac{\sum_k |c_{ik}| \Sigma^\eta_k }{\Sigma^q_{i} + \sum_{k} c_{ik}^2 \Sigma^\eta_{k}}\cdot (\tilde{a}_i - A_i) = \frac{\sum_k c_{ik}^2 \Sigma^\eta_k }{\Sigma^q_{i} + \sum_{k} c_{ik}^2 \Sigma^\eta_{k}}\cdot (\tilde{a}_i - A_i )
$$
Adding an irrelevant group increases the numerator and denominator on the right-hand side by the same amount. Since the numerator is smaller, the fraction increases. \qed

\bigskip

\noindent \textbf{Proof of Corollary \ref{cor:political correctness}.} Note that $c_{iK+1}^2 = 1$. For any member $j \neq i$ of group $\kappa$, $c_{iK+1} c_{jK+1} = 1$. Hence, adding group $K+1$ increases the numerator and denominator on the right-hand side of Equation \eqref{eq:biases_individuals} by the same amount. Since the ratio has absolute value less than 1, this increases the ratio. 

For any member $j$ of group $K+1$, $c_{iK+1}c_{jK+1} = -1$. Hence, adding group $K+1$ lowers the numerator on the right-hand side of Equation \eqref{eq:biases_individuals}, and raises the denominator by the same amount. Since the ratio has absolute value less than 1, this lowers the ratio. \qed

\bigskip

\noindent \textbf{Proof of Corollary \ref{cor:info about discrimination}.} Obvious from Equation \eqref{eq:biases_discrimination}.

\bigskip

\noindent \textbf{Proof of Corollary \ref{cor:competitor outsiders}.} The negative bias about members of group $K+1$ follows from the facts that for any $j>I$, $c_{iK+1}c_{jK+1} = -1$ and $c_{jk} = 0$ for any $k\leq K$. The second part follows from the fact that for any $j\leq I$, $c_{iK+1}c_{jK+1} = 1$, and that for $\Sigma^\eta_{K+1}$ sufficiently large, this term dominates. \qed

\bigskip

\noindent \textbf{Proof of Proposition \ref{prop:correlated_errors}.} We apply Part III Theorem \ref{thm:beliefs_no_action} to $f=a$, $M=Id$. Then, $[M^T \Sigma^{-1} M]= \Sigma$, and $M(\tilde f - f) = \tilde a - A$, yielding the formulas in the proposition. \qed

\bigskip

\noindent \textbf{Proof of Proposition \ref{prop:general_richer_obs}.}
Again the model is a special case of our general model introduced in Section \ref{sec:theoretical_tools} with 
\[
	\begin{pmatrix}
	q\\
	\eta\\
	s
	\end{pmatrix} = M 
	\begin{pmatrix}
	a\\
	\theta
	\end{pmatrix} + \epsilon \,,
\]
where $\epsilon \sim N(0,\Sigma)$. We have that the matrix $M$ is given 
\[
    M = \begin{pmatrix}
    Id & C \\
    0 & Id \\
    Id & 0 
    \end{pmatrix}
\]
and the variance covariance matrix is of the form
\[
    \Sigma = \begin{pmatrix}
    v^q Id & 0 & 0 \\
    0 & v^\eta Id & 0 \\
    0 & 0 &v^a Id
    \end{pmatrix}\,.
\]
By Theorem \ref{thm:beliefs_no_action} (III), we have that the agent's long-run bias is given by
\begin{align} \label{eq:bias-prop3}
	\Delta_k &= \frac{\left[M^{^T} \Sigma^{-1} M\right]^{-1}_{ki}}{\left[M^{^T} \Sigma^{-1} M\right]^{-1}_{ii} } \Delta_i\,.
\end{align}
To compute the agents beliefs we first compute $(M^T \Sigma^{-1} M)^{-1}$. We get that
\begin{align*}
    M^T \Sigma^{-1} M &= \begin{pmatrix}
    Id & 0 & Id\\
    C^T & Id & 0
    \end{pmatrix} \times 
    \begin{pmatrix}
    \frac{1}{v^q} Id & 0 & 0 \\
    0 & \frac{1}{v^\eta} Id & 0 \\
    0 & 0 &\frac{1}{v^a} Id
    \end{pmatrix} \times
    \begin{pmatrix}
    Id & C \\
    0 & Id \\
    Id & 0 
    \end{pmatrix}\\
    &= \begin{pmatrix}
    Id & 0 & Id\\
    C^T & Id & 0
    \end{pmatrix} \times
    \begin{pmatrix}
    \frac{1}{v^q} Id & \frac{1}{v^q} C\\
    0 & \frac{1}{v^\eta} Id \\
    \frac{1}{v^a} Id & 0 
    \end{pmatrix}\\
    &= \begin{pmatrix}
    \left( \frac{1}{v^q} + \frac{1}{v^a}\right) Id & \frac{1}{v^q} C\\
    \frac{1}{v^q} C^T & \frac{1}{v^\eta} Id + \frac{1}{v^q} C^T C
    \end{pmatrix}\,.
\end{align*}
The inverse to this matrix is given by
\begin{align*}
    &[M^T \Sigma^{-1} M]^{-1} = \begin{pmatrix}
        \left( \frac{1}{v^q} + \frac{1}{v^a}\right) \frac{1}{v^\eta} Id + \frac{1}{v^q v^a}  C C^T & 0 \\
        0 & \left( \frac{1}{v^q} + \frac{1}{v^a}\right) \frac{1}{v^\eta} Id + \frac{1}{v^q v^a}  C^T C
    \end{pmatrix}^{-1} \\
    & \hspace{5cm}\times \begin{pmatrix}
        \frac{1}{v^\eta} Id  + (C C^T) \frac{1}{v^q} & -C \frac{1}{v^q}\\
        -C^T \frac{1}{v^q} & \left( \frac{1}{v^q} + \frac{1}{v^a}\right) Id
    \end{pmatrix} \\
    &= \begin{pmatrix}
        \left[ \left( \frac{1}{v^q} + \frac{1}{v^a}\right) \frac{1}{v^\eta} Id + \frac{1}{v^q v^a}  C C^T \right] ^ {-1} \left[ \frac{1}{v^\eta} Id +  (C C^T) \frac{1}{v^q}\right] & - \left[ \left( \frac{1}{v^q} + \frac{1}{v^a}\right) \frac{1}{v^\eta} Id + \frac{1}{v^q v^a}  C C^T \right] ^ {-1} C \frac{1}{v^q}\\
        - \left[ \left( \frac{1}{v^q} + \frac{1}{v^a}\right) \frac{1}{v^\eta} Id + \frac{1}{v^q v^a}  C^T C\right] ^ {-1} C^T \frac{1}{v^q} & \left[ \left( \frac{1}{v^q} + \frac{1}{v^a}\right) \frac{1}{v^\eta} Id + \frac{1}{v^q v^a}  C^T C\right] ^ {-1} \left( \frac{1}{v^q} + \frac{1}{v^a}\right) Id
    \end{pmatrix} \,.
\end{align*}
To identify agent $i$'s biases regarding other individuals, we need to understand the upper left corner of this matrix. Furthermore, since each bias given in \eqref{eq:bias-prop3} is given by the ratio of two matrix entries, it is sufficient to understand the matrix up to a multiplicative constant. The matrix is proportional to 
\begin{equation*}
\left[ \frac{v^q+v^a}{v^\eta} Id + CC^T \right]^{-1} \left[ \frac{v^q}{v^\eta} Id + CC^T \right]. 
\end{equation*}
Define $x = \frac{v^q+v^a}{v^\eta} \in \RR $ and $y=\frac{v^q}{v^\eta} \in \RR$. Rewriting gives
\begin{equation}\label{eq:ability-bias}
\big[xId + CC^T\big]^{-1} \big[xId + CC^T+ (y-x)Id\big] = Id + (y-x) \big[xId + CC^T\big]^{-1} \,.
\end{equation}
We consider the special case in which there is one group, and each individual in the population is either a member or a competitor of the group. This means that $C$ is an $N$-dimensional vector consisting only of $+1$'s and $-1$'s. Notice that in this case 
\[
(CC^T)^2_{ij} = \sum_k (CC^T)_{ik}(CC^T)_{kj} = \sum_k c_i c_k^2 c_j = \sum_k c_ic_j = N \,(CC^T)_{ij}, 
\]
so that $(CC^T)^2 = I \, CC^T$. Given this, we have that 
\begin{align*}
	\big[x\,Id + CC^T\big] \left( \frac{1}{x} Id - \frac{1}{x^2 + I\,x}CC^T\right) &= Id - \frac{x}{x^2 + I\,x} C C^T + \frac{1}{x} CC^T - \frac{1}{x^2 + I\,x} CC^T CC^T\\
	&= Id - \frac{x}{x^2 + I\,x} C C^T + \frac{1}{x} CC^T - \frac{I}{x^2 + I\,x} CC^T \\
	& = Id \,,
\end{align*}
and thus
$$
\big[xId + CC^T\big]^{-1} = \frac{1}{x} Id - \frac{1}{x^2 + I\,x}CC^T \,.
$$
As a consequence we get that \eqref{eq:ability-bias} simplifies to
$$
Id + (y-x) \big[xId + CC^T\big]^{-1}  =Id + \frac{y-x}{x} Id - \frac{y-x}{x^2 + I\,x}CC^T = \frac{y}{x} Id + \frac{x-y}{x^2 + I\,x}CC^T.
$$
Plugging in for $x$ and $y$ yields 
$$
\frac{v^q}{ v^q+ v^a}Id + \frac{\frac{v^a}{v^\eta}}{\left(\frac{v^q+v^a}{v^\eta}\right)^2 + I\, \frac{v^q+v^a}{v^\eta} }CC^T,
$$
which is proportional to 
$$
v^q Id + \frac{v^a}{\frac{v^q+v^a}{v^\eta} + I}CC^T \,.
$$
Hence, agent $i$'s bias regarding agent $j$ satisfies
\begin{align*}
\frac{\tilde{a}^i_j - A_j}{\tilde{a}_i - A_i} &= \frac{\left( v^q Id + \frac{v^a}{\frac{v^q+v^a}{v^\eta} + I}CC^T \right)_{ij}}{\left( v^q Id + \frac{v^a}{\frac{v^q+v^a}{v^\eta} + I}CC^T \right)_{ii}} \\
&= \frac{\frac{v^a}{\frac{v^q+v^a}{v^\eta} + I}c_ic_j}{v^q + \frac{v^a}{\frac{v^q+v^a}{v^\eta} + I}} = \frac{v^{\eta} v^a c_i c_j}{(v^q +v^{\eta})(v^q+v^a) + (I-1)v^qv^{\eta}} \,.
\end{align*}
Similarly, the upper right corner of the matrix is proportional (with the same proportionality) to 
$$
-v^{\eta} C + \frac{v^{\eta}}{\frac{v^q+v^a}{v^{\eta} }+I }CC^TC
$$
Notice that $CC^TC = I\,C$. Hence, the $i$th component of the above vector equals 
$$
-\frac{v^q+ v^a}{\frac{v^q+v^a}{v^{\eta} }+I}c_i, 
$$
and therefore agent $i$'s bias about discrimination is 
$$
\frac{\tilde{\theta}^i_1 - \Theta_1}{\tilde{a}_i - A_i} = \frac{-\frac{v^q+ v^a}{\frac{v^q+v^a}{v^{\eta} }+I}c_i}{v^q + \frac{v^a}{\frac{v^q+v^a}{v^\eta} + I}}= \frac{-v^{\eta}(v^q+v^a)c_i}{(v^q +v^{\eta})(v^q+v^a) + (I-1)v^qv^{\eta} }.
$$

\bigskip

\noindent \textbf{Calculations behind Example \ref{prop:richer_observations}.} In the notation of Theorem \ref{thm:beliefs_no_action}, 
$$
f = \begin{pmatrix}
	a_1 \\ a_2 \\ a_3 \\ a_4 \\ \theta_1
\end{pmatrix}, \ \ 
M = \begin{pmatrix}
1 &	0	& 0 &	0 &	 1 \\	
0 &	1	& 0 &	0 &	 1	\\
0 &	0	& 1	& 0	& -1	 \\
0 &	0	& 0 & 	1 &	-1	\\
0 &	0 &	0 &	0 &	 1 \\	
1 &	0 &	0 &	0 &	 0 \\
0 &	1 &	0 &	0 &	 0 \\
0 &	0 &	1 &	0 &	 0 \\
0 &	0 &	0 &	1 &	 0
\end{pmatrix}, \ \
\Sigma = \begin{pmatrix}
	 1 &   0 &   0 &   0 &  0 &  0 &   0 &   0 &   0 \\
  0 &  1 &   0 &   0 &  0 &  0 &   0 &   0 &   0 \\
  0 &   0 &  \sigma^q_o &   0 &  0 &  0 &   0 &   0 &   0 \\
  0 &   0 &   0 &  \sigma^q_o &  0 &  0 &   0 &   0 &   0 \\
  0 &   0 &   0 &   0 &  1 &  0 &   0 &   0 &   0 \\
  0 &   0 &   0 &   0 &  0 & 1 &   0 &   0 &   0 \\
  0 &   0 &   0 &   0 &  0 &  0 &  1 &   0 &   0 \\
  0 &   0 &   0 &   0 &  0 &  0 &   0 &  \sigma^a_o &   0 \\
  0 &   0 &   0 &   0 &  0 &  0 &   0 &   0 &  \sigma^a_o 
\end{pmatrix}.
$$
Applying Part III of Theorem \ref{thm:beliefs_no_action}, and using Matlab in symbolic mode yields the results. \qed 

\bigskip

\noindent \textbf{Calculations behind Example \ref{ex:multi-dim_attr}.} Let $a_1' = a_1+m_1$ and $a_2' = a_2 + m_2$. In the notation of Theorem \ref{thm:beliefs_no_action}, 
$$
f = \begin{pmatrix}
	a_1' \\ a_2' \\ \theta_1 \\ a_2 
\end{pmatrix}, \ \ 
M = \begin{pmatrix}
1 &	0	& 1 &	0  \\	
0 &	1	& -1 &	0	\\
0 &	1	& 0	& 1 \\
0 &	0	& 1 & 0	
\end{pmatrix}, \ \
\Sigma = \begin{pmatrix}
	 v^q_1 &   0 &   0 &   0  \\
  0 &  v^q_2 &   0 &   0 &   \\
  0 &   0 &  v^b_2 &   0  \\
  0 &   0 &   0 &  v^{\eta}_1 
\end{pmatrix}.
$$
Applying Part III of Theorem \ref{thm:beliefs_no_action}, and using Matlab in symbolic mode yields the results. \qed

\end{appendix}

\end{document}